\documentclass[letterpaper,11pt]{article}
\pdfoutput=1

\usepackage{jheppub}
\usepackage{multirow}

\usepackage{subfig}
\usepackage{xspace}
\usepackage[countmax]{subfloat}

\usepackage{amssymb}
\usepackage{amsmath}
\usepackage{cancel}
\usepackage{color}
\usepackage{braket}
\usepackage{graphicx}
\usepackage{csquotes}
\usepackage{multirow}
\usepackage{verbatim}
\usepackage{amsthm}
\usepackage{slashed}
\usepackage{wasysym}
\usepackage{simplewick}
\usepackage{mathtools}
\usepackage{soul}
\usepackage{xspace}
\usepackage{microtype}
\usepackage[separate-uncertainty = true,multi-part-units=single]{siunitx}
\usepackage{cleveref}
\usepackage[parfill]{parskip}

\usepackage{scalefnt}
\newcommand{\abbrev}{\scalefont{.9}}

\newcommand{\Tau}{\mathcal{T}}

\newcount\colveccount
\newcommand*\colvec[1]{
        \global\colveccount#1
        \begin{pmatrix}
        \colvecnext
}
\def\colvecnext#1{
        #1
        \global\advance\colveccount-1
        \ifnum\colveccount>0
                \\
                \expandafter\colvecnext
        \else
                \end{pmatrix}
        \fi
}

\newcommand{\refcite}[1]{ref.~\cite{#1}}
\newcommand{\refscite}[1]{refs.~\cite{#1}}
\newcommand{\eq}[1]{eq.~\eqref{eq:#1}}

\renewcommand{\sec}[1]{sec.~\ref{sec:#1}}

\newcommand{\tab}[1]{Table~\ref{tab:#1}}
\newcommand{\fig}[1]{fig.~\ref{fig:#1}}

\def\cO{\mathcal{O}}

\def\nn{{\nonumber}}

\def\be{\begin{equation}}
\def\ee{\end{equation}}

\newcommand{\tcut}{{\tau_{\text{cut}}}}
\newcommand{\tauc}{\tcut}
\newcommand{\Tcut}{{\Tau_{0\,\text{cut}}}}
\newcommand{\Tauc}{\Tcut}

\newcommand{\robs}{x}
\newcommand{\rcutnobrack}{\robs_{\text{cut}}}
\newcommand{\rcut}{{\robs_{\text{cut}}}}

\newcommand{\dsigmadrobs}{\frac{\df \sigma}{\df \robs}} 
 
\newcommand{\dsigmadrobsat}[1]{\frac{\df \sigma^{#1}}{\df \robs}}

\allowdisplaybreaks[3]

\newcommand{\df}{\mathrm{d}}

\newcommand\bn{{\bar n}}

\newcommand{\lep}{\mathrm{lep}}
\newcommand{\hadcm}{\mathrm{cm}}

\newcommand{\NLO}{\text{\abbrev NLO}}
\newcommand{\NNLO}{\text{\abbrev NNLO}}
\newcommand{\NNNLO}{\text{\abbrev N$^3$LO}}
\newcommand{\MCFM}{\text{\abbrev MCFM}}
\newcommand{\MATRIX}{{\sc Matrix}\xspace}
\newcommand{\QCD}{\text{\abbrev QCD}}
\newcommand{\LHC}{\text{\abbrev LHC}}
\newcommand{\PTB}{\text{\abbrev P2B}}
\newcommand{\PDF}{\text{\abbrev PDF}}

\newcommand{\iso}{\mathrm{iso}}
\newcommand{\Obs}{\cO}
\newcommand{\Obst}{\tilde\cO}

\newcommand{\GeV}{\:\mathrm{GeV}}

\newcommand{\Ecm}{E_\mathrm{cm}}

\newcommand{\mathrmNnminusoneLO}{{\mathrm{N^{n-1}LO}}}
\newcommand{\mathrmNnLO}{{\mathrm{N^{n}LO}}}

\newcommand{\PtoBqt}{{\abbrev P2B}-$q_T$ subtraction}
\newcommand{\PtoBtau}{{\abbrev P2B}-$\tau_0$ subtraction}
\newcommand{\PtoBtauandqt}{{\abbrev P2B}-$\tau_0$ and {\abbrev P2B}-$q_T$ subtractions}

\newcommand{\ndelta}{n_{\delta}}
\newcommand{\bndelta}{{\bar{n}_{\delta}}}

\newcommand{\blue}[1]{{\color{blue}#1}}
\newcommand{\red}[1]{{\color{red}#1}}

\newcommand{\inclcolor}[1]{{\color[rgb]{0,.6,0}#1}}
\newcommand{\kinematiccolor}[1]{\red{#1}}
\newcommand{\isocolor}[1]{\blue{#1}}

\newcounter{notecount}

  \newcount\hour \newcount\minute
  \hour=\time \divide \hour by 60 \minute=\time
  \newcommand{\todaytime}{\today \ -- \number\hour :\ifnum \minute<10 0\fi\number\minute}

\preprint{CERN-TH-2024-102, FERMILAB-PUB-24-0402-T}
\title{Projection-to-Born-improved Subtractions at NNLO}

\abstract{
While the current frontier in fixed-order precision for collider observables is \NNNLO{}, important steps are necessary to 
consolidate \NNLO{} cross-section predictions with improved stability and efficiency.
Slicing methods have been successfully applied to obtain \NNLO{} and \NNNLO{} predictions, but have 
shown poor performance in the presence of fiducial cuts due to large kinematical power corrections.
In this paper we implement Projection-to-Born-improved $q_T$ (\PTB{} $q_T$) and jettiness (\PTB{} 
$\tau_0$) subtractions for a large 
class of color singlet processes in \MCFM{}. 
This method allows for the efficient evaluation of \emph{fiducial} power corrections in any 
non-local 
subtraction scheme using a Projection-to-Born subtraction.
We demonstrate the significant numerical improvements of this method based on fiducial Drell-Yan 
and Higgs 
cross-sections.
Moreover, with fiducial power corrections removed via this method, the leading-logarithmic 
power corrections that have only been calculated without fiducial cuts can be included, further 
improving the calculations.
For di-photon production with photon isolation, we devise a novel method in combination with 
\PTB{}-improved subtractions, which we name $\PTB{}_\gamma$ $\tau_0$, and $\PTB{}_\gamma ~q_T$  
for the two subtraction schemes, respectively. 
This method allows the inclusion of both fiducial power corrections due to kinematic cuts on the 
photons and a set of isolation power corrections in the fragmentation channel where a quark may 
enter the isolation cone.
We find significant improvements in the convergence of \NNLO{} di-photon cross-sections with photon 
isolation cuts, demonstrating that it is possible to achieve a stable and efficient calculation 
of di-photon cross-sections using slicing methods. 
}

\begin{document}

\author[1]{John Campbell,}
\author[2]{Tobias Neumann,}
\author[3]{and Gherardo Vita}
\emailAdd{johnmc@fnal.gov}
\emailAdd{tneumann@smu.edu}
\emailAdd{gherardo.vita@cern.ch}

\affiliation[1]{Fermilab, PO Box 500, Batavia, Illinois 60510, USA}	
\affiliation[2]{Department of Physics, William and Mary, Williamsburg, Virginia, USA}
\affiliation[3]{CERN, Theoretical Physics Department, CH-1211 Geneva 23, Switzerland}

\maketitle

\section{Introduction}\label{sec:intro}
The Large Hadron Collider (\LHC{}) has been instrumental in advancing our understanding of 
fundamental physics, largely due to its capability to measure Standard Model (SM) processes with 
unprecedented precision.
\LHC{} measurements of key benchmark processes, such as the production of 
electroweak bosons ($W^\pm$, $Z$), have reached an astonishing level of accuracy, previously only 
limited by the luminosity uncertainty	 \cite{CMS:2021xjt,ATLAS:2022hro}, see for example 
refs.~\cite{ATLAS:2024nrd,CMS:2019raw,ATLAS:2019zci}.
Such a level of precision has not yet been reached for Higgs boson measurements since the main source of 
experimental uncertainty comes from the limited statistical power of current data sets 
\cite{ATLAS:2022fnp,ATLAS:2022qef,CMS:2023gjz,ATLAS:2023tnc}. 
However, the situation will be dramatically different for a projected seven-fold increase of 
statistics at the High Luminosity \LHC{} ({\abbrev 
HL-LHC}) \cite{ATLAS:2019mfr,ATLAS:2022hsp}.

In order to match the accuracy of data from the \LHC{},
precise theoretical predictions for Standard Model 
scattering processes are required.
To this end, one must include radiative corrections from quantum 
chromodynamics (\QCD{}) at least to the next-to-next-to-leading order (\NNLO{}), and, crucially, 
incorporate experimental cuts employed to define the fiducial region.  These are typically bounds on the 
transverse momenta and rapidity of final state leptons, or photon isolation cuts. Certain 
cuts used in experimental measurements are known to induce numerical instabilities or make the 
calculation more challenging due to infrared ({\abbrev IR}) sensitivities 
\cite{Frixione:1997ks,Salam:2021tbm}. 

The calculation of perturbative higher-order corrections requires the isolation and cancellation of
infrared singularities between real and virtual contributions, which
proceeds through so-called subtraction methods. There are two main 
classes of methods for performing
this procedure: infrared singularities can either be subtracted after integration over the
whole phase space (slicing  methods), or through the construction of more local, or
point-wise (fully local) counterterms. 

While fully differential \NNNLO{} predictions for hadronic collisions are the current frontier in 
fixed-order 
precision, reached for only a few processes 
\cite{Chen:2021isd,Billis:2021ecs,Chen:2021vtu,Chen:2022cgv,Chen:2022lwc,Neumann:2022lft,Campbell:2023lcy},
slicing 
and local subtraction 
methods for \NNLO{} cross-sections are moving towards full process 
generalization, consolidation and increased computational efficiency, see e.g. 
refs.~\cite{Devoto:2023rpv,Chen:2022ktf,Bertolotti:2022aih,TorresBobadilla:2020ekr} and references 
therein. Slicing methods are often easier to derive and implement, but can be numerically more 
challenging due to non-local cancellations compared to local subtractions. On the other 
hand, local subtractions require careful handling of all possible singular configurations and as a result
can suffer from a proliferation of subtraction terms with various consequences for numerical 
efficiency.

While local subtractions are naively numerically superior as singularities cancel for each 
integration point, a feature of slicing methods is that they can be systematically improved by 
computing perturbative power corrections to the leading-power factorization formulae upon which 
they rely. 
In the last decade, significant progress has been achieved in understanding collider observables beyond leading power
\cite{Moult:2016fqy,Boughezal:2016zws,Moult:2017jsg,Boughezal:2018mvf,	
Ebert:2018lzn,Ebert:2018gsn,Boughezal:2019ggi,Moult:2019uhz,Moult:2018jjd,Beneke:2019mua,Liu:2019oav,Liu:2020tzd,Beneke:2020ibj,vanBeekveld:2021hhv,Beneke:2022obx,Liu:2022ajh,Ferrera:2023vsw,Vita:2024ypr,Beneke:2024cpq},
making the calculation of power corrections for slicing methods an obvious application of this 
program.
The inclusion of such power corrections in slicing methods offers the possibility of leveling or 
even tilting the playing field, by
increasing numerical stability and performance by orders of magnitude. 
However, the interplay between the subtraction of {\abbrev IR} singularities and the phase space 
cuts induced by the definition of a fiducial region requires particular attention and can spoil the 
expected performance of these methods.

A particularly efficient local subtraction method is the so-called Projection-to-Born 
(\PTB{}) 
scheme \cite{Cacciari:2015jma}, 
which has recently been extended to the calculation of fully differential Higgs production at the 
\LHC{} \cite{Chen:2021isd}.
The numerical advantage of \PTB{} stems from using actual process matrix elements as subtraction 
counterterms, obtained by projecting the full phase space onto Born configurations.
The drawback of this approach lies in the complexity of obtaining the \emph{integrated} 
counterterms, as this consists of the exact amplitudes integrated over the projected phase space. 
Recently, it has been shown that fiducial and hadronic power corrections in slicing methods
can be disentangled and that the former can be efficiently evaluated using a Projection-to-Born (\PTB{}) prescription 
\cite{Ebert:2019zkb,Vita:2024ypr}.
This method can be applied to both $q_T$ and $0$-jettiness subtractions, and we refer to it as 
Projection-to-Born-improved $q_T$ subtraction (\PtoBqt) and Projection-to-Born-improved $\tau_0$ 
subtraction (\PtoBtau).

In this paper we implement \PTB{}-improved subtractions in the public code \MCFM{}. 
We discuss the improvements in terms of stability and efficiency for the calculation of \NNLO{} 
cross-sections in the presence of fiducial cuts.
In \cref{sec:setup} we review \PTB{}-improved non-local subtractions and provide some details of 
the impact of various components at \NLO{}.
In \cref{sec:DY,sec:Higgs} we study fiducial \NNLO{} cross-sections using \PtoBtau{} 
for di-lepton $Z$ production and Higgs production in gluon fusion.
In \cref{sec:diphoton}, we study power corrections in di-photon production at \NNLO{}. 
The di-photon process is particular challenging since its measurement crucially relies on photon isolation 
cuts, which are known to cause severe numerical instabilities at higher orders in \QCD{}. 
For this case, we introduce a novel method to include a sizable set of isolation power corrections that make it possible 
to obtain numerically reliable results. These additional power corrections cannot be 
accounted for by a recoil prescription~\cite{Ebert:2019zkb,Ebert:2020dfc,Catani:2015vma} in the case of $q_T$
subtractions.
We conclude in \cref{sec:conc}.  

\section{Projection-to-Born improved $0$-jettiness and $q_T$ subtractions}
\label{sec:setup}
We first set up the notation for non-local subtractions, discuss the general setup for 
\PtoBtauandqt{} and present a general classification of power corrections in jettiness and $q_T$ 
subtractions in terms of hadronic and fiducial components.

\subsection{Review of $0$-jettiness and $q_T$ subtractions}

Jettiness subtractions \cite{Boughezal:2015dva,Gaunt:2015pea} and $q_T$ subtractions 
\cite{Catani:2007vq} differ in the observable used for identifying and subtracting 
infrared singularities. 
For $q_T$ subtractions, the transverse momentum $q_T$ is simply defined as the Euclidean norm of 
the transverse momentum 
components of the color-singlet system momentum.
For $0$-jettiness the observable is defined by \cite{Stewart:2010tn}
\begin{align} \label{eq:Tau0_0}
 \Tau_0 = \sum_i \min \biggl\{ \frac{2 p_a \cdot k_i}{Q_a} \,, \frac{2 p_b \cdot k_i}{Q_b} \biggr\}
\,,\end{align}
where the sum runs over the final state momenta of the color-charged particles with momenta $k_i$. 
The reference incoming Born momenta $p_{a,b}^\mu$ are defined as 
\begin{align} \label{eq:Born}
 p_a^\mu = x_a \Ecm \frac{n^\mu}{2} = Q e^Y \frac{n^\mu}{2}
\,,\qquad
 p_b^\mu = x_b \Ecm \frac{\bn^\mu}{2} = Q e^{-Y} \frac{\bn^\mu}{2}
\,,\end{align}
and we identified with $Y$ and $Q$ the color singlet rapidity and invariant mass.
We define the lower-case observable $\tau_0$ through the dimensionless ratio $\tau_0=\Tau_0/Q$, following 
the convention often found in the literature.

The 0-jettiness definition of \eq{Tau0_0} allows for some freedom in the choice of the normalization factors
$Q_{a,b}$ and fixing these normalizations leads to definitions that differ only beyond leading power.
In particular, we have two common choices,
\begin{alignat}{4} \label{eq:Tau0_2}
 &\text{leptonic:}\quad & Q_a &= Q_b = Q \,,\qquad
  & \Tau_0^\lep
  &= \sum_i \min \biggl\{ \frac{x_a \Ecm}{Q} n \cdot k_i \,,\, \frac{x_b \Ecm}{Q} \bn \cdot k_i \biggr\}
\nn\\* &&&&
  &= \sum_i \min \biggl\{ e^Y n \cdot k_i \,,\, e^{-Y} \bn \cdot k_i \biggr\}
\nn\\
 &\text{hadronic:}\qquad & Q_{a,b} &= x_{a,b} \Ecm \,,
 & \Tau_0^\hadcm &= \sum_i \min \Bigl\{ n \cdot k_i \,,\, \bn \cdot k_i \Bigr\}
\,.\end{alignat}
For $0$-jettiness subtractions \cite{Boughezal:2015dva,Gaunt:2015pea}, to be introduced below, the 
leptonic definition has 
significantly 
smaller power corrections thanks to its invariance under 
longitudinal boosts of the frame in which the minimization is performed 
\cite{Moult:2016fqy,Moult:2017jsg,Ebert:2018lzn}. 
The numerical implications of this phenomenon in \MCFM{} have been studied in depth in 
\refcite{Campbell:2019dru} with the release of \MCFM{}~9, and since then the leptonic definition 
has been adopted as the default choice. 
In the remainder of this paper we therefore consider only the leptonic definition of $\Tau_0$.

\paragraph{Slicing subtractions based on $q_T$ and $0$-jettiness.}
In this paper we consider the calculation of cross-sections $\sigma$ for an observable $\cO$ using 
a slicing variable $\robs$, where we focus on $\robs=\tau_0=\Tau_0/Q$ ($0$-jettiness subtractions) 
\cite{Boughezal:2015dva,Gaunt:2015pea}
and $\robs=q_T^2/Q^2$ ($q_T$~subtractions) 
\cite{Catani:2007vq}.  
Similarly, we denote the cutoff variables as 
\be
\rcut = \Tau_{0,\,\mathrm{cut}}/Q \equiv \tauc \,,\qquad\rcut = q_{T\mathrm{cut}}^2/Q^2\,.
\ee 
The subtraction method is organized as follows.
\begin{align}\label{eq:slicing}
	\sigma(\cO) &= \int_0^\rcut \df \robs \dsigmadrobs(\cO) +  \int_\rcut^{\robs_\text{max}} \df 
	\robs \dsigmadrobs(\cO) 
	\nn\\ 
	 &= \int_0^{\rcut} \df \robs \dsigmadrobsat{\text{sub}} (\cO) + \int_0^{\rcut} \df \robs 
	\left[  \dsigmadrobs - \dsigmadrobsat{\text{sub}} \right](\cO) + \int_\rcut^{\robs_\text{max}} 
	\df \robs \dsigmadrobs (\cO) \,.
	\nn\\ &\equiv \sigma_\text{sub}(\rcut,\cO) + \Delta\sigma(\rcut,\cO) + 
	\int_\rcut^{\robs_\text{max}} \df \robs \dsigmadrobs (\cO) \,.
\end{align}  
The observables $\robs$ are designed to regularize the infrared singularities with a cutoff 
$\rcut$, 
and 
the cross-section is split into parts $\robs>\rcut$ and $\robs<\rcut$. For $\robs<\rcut$ a 
factorization 
theorem is used to describe \QCD{} in the region of soft and collinear kinematics, while, by 
construction, $\robs>\rcut$ consists of the process under consideration with additional \QCD{} 
radiation at a lower order. 

For the discussion in this paper we do not need a detailed description of the below-cut part that 
also further depends in detail on the observable $\robs$. 
However, it is important to stress that the leading-power behavior of these observables is very well understood. 
For $q_T$-subtractions, the leading-power factorization theorem was originally established in 
\refscite{Collins:1981uk,Collins:1981va,Collins:1984kg} and subsequently revised in various 
formalisms throughout the years 
\cite{Catani:2000vq,deFlorian:2001zd,Catani:2010pd,Collins:1350496,Becher:2010tm,Becher:2011xn,Becher:2012yn,GarciaEchevarria:2011rb,Echevarria:2012js,
 Echevarria:2014rua,Chiu:2012ir,Li:2016axz}. 
With the recent calculation of the three-loop $q_T$ beam functions 
\cite{Luo:2019szz,Ebert:2020yqt}, and the previously known color-singlet hard functions \cite{Gehrmann:2010ue} and 
$q_T$ soft function \cite{Li:2016ctv}, the fixed-order expansion at \NNNLO{} at leading power 
({\abbrev LP}) is now 
fully 
available.
This has paved the way for the first \NNNLO{} predictions for fully differential color-singlet 
production using the $q_T$ subtraction method 
\cite{Billis:2021ecs,Chen:2021vtu,Chen:2022cgv,Chen:2022lwc,Neumann:2022lft,Campbell:2023lcy}.
For $N$-jettiness, the leading-power factorization was given in \refcite{Stewart:2010tn} using 
Soft-Collinear Effective Theory ({\abbrev SCET})
\cite{Bauer:2000ew, Bauer:2000yr, Bauer:2001ct, Bauer:2001yt}. 
The \NNLO{} fixed-order expansion of all the objects entering the factorization are available 
\cite{Gaunt:2014cfa,Gaunt:2014xga,Boughezal:2017tdd,Kelley:2011ng,Monni:2011gb,Boughezal:2015eha,Li:2016tvb,Campbell:2017hsw,Alioli:2021ggd,Bell:2023yso,Agarwal:2024gws},
 and in the past years significant progress has been made in extending these calculations to 
\NNNLO{} 
\cite{Bruser:2018rad,Ebert:2020lxs,Ebert:2020unb,Baranowski:2021gxe,Chen:2020dpk,Baranowski:2024ene}.

In \cref{eq:slicing} we define a 
subtracted term (usually obtained by deriving a factorization formula in $\robs$)
\be
	\sigma_\text{sub}(\rcut,\cO) \equiv \int_0^{\rcut} \df \robs \dsigmadrobsat{\text{sub}}(\cO)\,,
\ee
for which we need to have analytic control as it encodes the singular behavior of the cross-section 
as the slicing variable $\rcut$ goes to $0$. We furthermore defined a slicing residual term
\be
	\Delta\sigma(\rcut,\cO) \equiv \int_0^{\rcut} \df \robs \left[  \dsigmadrobs - 
	\dsigmadrobsat{\text{sub}} \right](\cO)\,,
\label{eq:residual}
\ee
that is determined by integrable terms over which we lack analytic control and are therefore compelled to neglect. 
This neglect constitutes the primary source of numerical uncertainty in the slicing procedure, and 
we refer to it as the slicing residual, slicing cutoff error, or error due to neglected power 
corrections. 
In the following section we review the analytic properties of these power corrections.

\subsection{Hadronic and fiducial power corrections in $0$-jettiness and $q_T$ subtractions}
\label{sec:fiducialpc}

To discuss power corrections, we start with the cumulant with respect to a slicing cutoff $\rcut$ 
for a given observable $\cO$. Order by order in perturbation theory it takes the form
\be\label{eq:xsscalinggeneral}
	\df\sigma(\rcut, \cO) \sim \sum_l \left(\frac{\alpha_s}{4\pi}\right)^l \left[ \sum_{m=0}^{2l} 
	c^{(0)}_{l,m}(\cO) \ln^m \rcut + \sum_j c^{(p)}_{l,j}(\cO) \rcut^{p}\ln^j \rcut  + \dots \right]
\ee
for some $p>0$, and the dots indicate terms that are further power suppressed as $\rcut\to 0$.
The first term in the bracket is divergent as $\rcut\to 0$ and must be included in the subtraction
term. We recognize it as the cumulant of the leading-power distribution
\be
	\df\sigma_\mathrm{LP}(\rcut) = \int_0^{\rcut} \df \robs \, \dsigmadrobsat{\text{LP}} \sim \sum_l \left(\frac{\alpha_s}{4\pi}\right)^l  \sum_{m=0}^{2l} c^{(0)}_{l,m} \ln^m \rcut \,.
\ee
The simplest and most common choice for slicing subtractions is to take the cumulant of the 
leading-power terms. 
Therefore, only the fixed-order expansion of the leading-power ({\abbrev LP}) factorization theorem 
is needed analytically.

For the residual subleading-power terms no factorization theorems exist so far. However, at fixed 
order they take the general form
\be \label{eq:deltasigmapscaling}
	\Delta\sigma(\rcut) \sim \rcutnobrack^{p} \sum_{j} \ln^j \rcut +\dots\,.
\ee
where $p$ crucially depends on the measurement (observable) constraints and, to a lesser degree, on 
the slicing variable. 
In the following we categorize these power corrections and discuss their scaling behavior.
\paragraph{Hadronic power corrections.}  
In the case of observables without fiducial cuts, we have $p=1$ \cite{Farhi:1977sg,Ellis:1981hk,Gaunt:2015pea,Moult:2016fqy,Moult:2017jsg,Ebert:2018lzn,Ebert:2018gsn,Ferrera:2023vsw}.
We refer to this class of perturbative power corrections as \emph{hadronic} or \emph{dynamical} power corrections, to distinguish them from the fiducial (and photon-isolation) power corrections 
that have an origin closely related to the kinematics of the non-\QCD{} interacting lepton or photon final states.
The hadronic power corrections can be written as
\be \label{eq:deltasigmahad}
	\Delta\sigma^{\rm had}(\rcut,\cO) \sim \rcut \left( c^{(1)}_{l,2l-1}(\cO) \ln^{2l -1} \rcut + c^{(1)}_{l,2l-2}(\cO) \ln^{2l -2} \rcut +\dots\right)\,,
\ee
where $l$ is the perturbative order, as in \eq{xsscalinggeneral}.
In this case, significant effort has been made to obtain analytic control of the first term in 
$\Delta\sigma^{\rm had}(\rcut)$. 
While this involves understanding and performing complicated next-to-leading-power calculations, 
results have been obtained for different processes and slicing schemes 
\cite{Moult:2016fqy,Boughezal:2016zws,Moult:2017jsg,Boughezal:2018mvf,	
Ebert:2018lzn,Ebert:2018gsn,Boughezal:2019ggi,Moult:2019uhz,Ferrera:2023vsw},
	 recently up to \NNNLO~\cite{Vita:2024ypr}.
	 
Power corrections from refs.~\cite{Moult:2016fqy,Moult:2017jsg,Ebert:2018lzn} were implemented in 
\MCFM{} in refs.~\cite{Campbell:2019dru,Campbell:2017aul} and studied extensively for total 
inclusive cross-sections and differential cross-sections in the absence of fiducial cuts.
In this paper we include those leading logarithmic ({\abbrev LL}) next-to-leading-power ({\abbrev 
NLP}) corrections in numerical 
comparisons with the label \enquote{{\abbrev NLP LL}}.

Note that the power $p=1$ for these hadronic power corrections has only been demonstrated for the case of a Drell-Yan like process with one hard scale $Q$. 
This may no longer hold for (color-singlet) multi-boson processes with more complicated leading-order topologies and therefore multiple hard scales.

\paragraph{Fiducial power corrections.}
While hadronic power corrections are important, fiducial 
cuts are always present in experimental measurements. These fiducial cuts induce additional power 
corrections that are therefore 
present in virtually every relevant theoretical prediction. 
Their impact on numerical calculations can be significant and understanding how they affect 
the 
behavior of the residual slicing error is crucial in obtaining efficient, stable and reliable 
results.

In \refcite{Ebert:2019zkb}, a first analytic examination of the structure of fiducial power corrections 
was performed: For a Drell-Yan like process with symmetric transverse momentum cuts on the leptons 
$p_T^l > p_T^\text{min}$ one has power corrections with $p=1/2$, proportional to 
$p_T^\text{min}/Q$. There is a further logarithmic suppression compared to the hadronic power 
corrections.
\be\label{eq:deltasigmafiducial}
	\Delta\sigma^{\mathrm{fiducial}}_{p_{T}^\ell}(\rcut) \sim \sqrt{\rcut}\ln^{2l-2}(\rcut) + 
	\dots\,.
\ee
Since these power corrections originate from breaking the azimuthal symmetry that is only present 
in the Born process, it is expected that $p=1/2$ for generic fiducial cuts 
\cite{Ebert:2019zkb}.

\subsection{Numerical calculation of fiducial power corrections using Projection-to-Born}
\label{sec:P2Bpc}
As we have analyzed in \sec{fiducialpc}, fiducial power corrections have a numerical impact that is often dominant over hadronic ones. 
Unfortunately, it is in general complicated to capture them analytically. However, one can account for them efficiently 
in a numerical manner by using a Projection-to-Born prescription \cite{Ebert:2019zkb,Vita:2024ypr}.
In short, for the calculation of an observable $\Obs$, point by point in the real emission phase 
space one calculates a Born projection of the momenta. The projected observable $\Obst$ is 
naturally calculated as the observable on this projected Born phase space. With $\Obs$ and $\Obst$ 
at hand one can calculate $\sigma(\Obs)$ in the following way,
\begin{align}
    \sigma_{h,\,\mathrmNnLO}(\Obs) &= \sigma_{h,\,\mathrmNnLO}(\Obst) + \sigma_{h+j,\,\mathrmNnminusoneLO}(\Obs - \Obst) \\[.5cm]
    &= \int^\rcut_0 \df \robs \frac{\df \sigma^\text{sub}_{h,\,\mathrmNnLO}}{\df \robs}(\Obst) + \int_{\robs > \rcut} \df \sigma^\text{full}_{h+j,\,\mathrmNnminusoneLO}(\Obst) &\text{\small(Slicing for $\Obst$)} \nonumber\\[.5cm]
    &\quad+  \int^\rcut_0 \df \robs \left[\frac{\df \sigma^\text{full}_{h,\,\mathrmNnLO}}{\df \robs} - \frac{\df \sigma^\text{sub}_{h,\,\mathrmNnLO}}{\df \robs}\right](\Obst)    &\parbox{4cm}{\small (Slicing residual $\Delta\sigma(\rcut)$ for $\Obst$. No dependence on fiducial cuts.)} \nonumber\\[.5cm] 
    &\quad+   \int \df \sigma^\text{full}_{h+j,\,\mathrmNnminusoneLO}(\Obs - \Obst)\,.    
    &\parbox{4cm}{\small (P2B correction factor for $\Obs$ vs $\Obst$)}\nn\
\end{align}
The slicing residual $\Delta\sigma$ does not depend on the fiducial cuts and can 
therefore be subject only to the hadronic power corrections of \eq{deltasigmahad}.
Note that the above-the-cut contributions may be combined, such that the \PTB{} corrections are 
calculated only below the cut,
\begin{align}
    \sigma_{h,\,\mathrmNnLO}(\Obs)   &= \int^\rcut_0 \df \robs \frac{\df 
    \sigma^\text{sub}_{h,\,\mathrmNnLO}}{\df \robs}(\Obs)  + \int_{\robs > \rcut} \df 
    \sigma^\text{full}_{h+j,\,\mathrmNnminusoneLO}(\Obs)\\[.5cm]
    &\quad+  \int^\rcut_0 \df \robs \left[\frac{\df \sigma^\text{full}_{h,\,\mathrmNnLO}}{\df \robs} - \frac{\df \sigma^\text{sub}_{h,\,\mathrmNnLO}}{\df \robs}\right](\Obst)   &\parbox{4cm}{\small (Slicing residual $\Delta\sigma(\rcut)$ for $\Obst$. No dependence on fiducial cuts.)} \nonumber\\[.5cm] 
    &\quad+   \int_0^\rcut \df \sigma^\text{full}_{h+j,\,\mathrmNnminusoneLO}(\Obs - \Obst)\,.    
    &\parbox{4cm}{\small (P2B correction factor for $\Obs$ vs $\Obst$, only below $\rcut$)} \nn
\end{align}

The \PTB{} method laid out here is a general way to compute the fiducial power corrections using an 
arbitrary slicing variable $x$. In the case of $x=q_T^2/Q^2$ an efficient alternative is to 
recoil-boost the Born kinematics \cite{Ebert:2019zkb,Ebert:2020dfc,Catani:2015vma} in the 
factorization formula to a finite value of $q_T$ and 
integrate up to the slicing cutoff $q_{T,\text{cut}}$. This prescription has been implemented in 
various codes \cite{Buonocore:2021tke,Camarda:2021jsw,Neumann:2022lft}. While the methods therefore 
seem equivalent for $q_T$, an important difference emerges in the presence of photon isolation. In 
this case the $q_T$ recoil cannot capture the associated power corrections since it starts from 
Born-level kinematics, while the \PTB{} method makes this possible. We discuss this in detail 
in \cref{sec:diphoton}.

\paragraph{Projection mapping.}
A crucial element of \PTB{}-improved subtractions is the concept of Born-projection of the partonic 
momenta.  
For our purposes, the \PTB{} map is a function that takes $3 + n$ momenta 
$\{p_a,p_b,q,k_1,\dots,k_n\}$ representing a $2 \to 1 + n$ scattering event and maps it to a set of 
3 momenta $\{\tilde p_a,\tilde p_b,\tilde q\}$ in a $2 \to 1$ scattering configuration. 
Following \refcite{Chen:2021isd},  we can obtain the projection-to-Born map via a redefinition of 
the 
incoming parton momenta
\begin{align}
 p_a \to \tilde p_a = \xi_a p_a\,,\qquad  p_b \to \tilde p_b = \xi_b p_b\,.
\end{align}
The map is completely fixed by the value of $\xi_{a,b}$ and by requiring momentum conservation for the projected momenta, i.e. $\tilde p_a^\mu + \tilde p_b^\mu= \tilde q^\mu$.
We can fix the fractions $\xi_{a,b}$ by imposing that the following relations hold,
\be
	q^2 = \tilde q^2\,,\qquad \frac{ \bn \cdot q }{n \cdot q} = \frac{ \bn \cdot \tilde q }{n \cdot \tilde q}\,,
\ee
which are trivially related to the preservation of the invariant mass and rapidity of the color singlet under the map. 
The $n$ and $\overline{n}$ vectors are normalized light-like directions, 
e.g. $n=p_a/|p_a|, \overline{n}=p_b/|p_b|$.
In the case of a subsequent decay of the color singlet to $j$ final state particles with momenta $\{q_1^\mu,\dots,q_j^\mu\}$ we can project their momenta as
\be
	q_i^\mu \to \tilde q_i^\mu = q_i^\mu + \frac{2 q_i \cdot q}{q^2} \tilde q^\mu - \frac{2(q + \tilde q) \cdot q_i}{(q + \tilde q)^2} (q + \tilde q)^\mu  \,,
\ee 
which accounts for the change of the rest frame of the color singlet after the \PTB{}.

Alternative mappings for the final state particles are possible. 
In fact, the one that we use for the predictions in  this paper is defined by considering a pure 
boost
\be\label{eq:lcprojection}
	q_i^\mu \to \tilde q_i^\mu = q_i^\mu + \left(\frac{\bndelta \cdot \tilde q}{\bndelta \cdot q} -1\right) \bndelta \cdot q_i \frac{\ndelta^\mu}{2}  + \left(\frac{\ndelta \cdot \tilde q}{\ndelta \cdot q} -1\right) \ndelta \cdot q_i \frac{\bndelta^\mu}{2}  \,,
\ee 
where 
\be
	\ndelta^\mu \equiv \left (1, \frac{\vec{\delta q}}{|\vec{\delta q}|}\right) \,,\qquad
	\bndelta^\mu \equiv  \left (1, -\frac{\vec{\delta q}}{|\vec{\delta q}|}\right)\,,\qquad \vec{\delta q} = \vec{\tilde q} - \vec{q} \,.
\ee
\section{Di-lepton and Higgs production at \NNLO{} with \PTB{}-improved subtractions}
\label{sec:DYLikenumerics}
We now study the numerical impact of \PTB{}-improved subtractions. We discuss $Z$ production 
with symmetric transverse momentum cuts in \cref{sec:DY} 
and gluon fusion Higgs production in \cref{sec:Higgs}.
In both cases we show results for a center-of-mass energy $\sqrt{s}=\SI{13}{\GeV}$ with the 
\PDF{} set \texttt{NNPDF31\_nnlo\_as\_0118} \cite{NNPDF:2017mvq}. The factorization and 
renormalization scales are set to the invariant mass $Q$ of the color-singlet system.

\subsection{$Z \to \ell^+ \ell^-$}
\label{sec:DY}

For $Z$ production we consider the case of symmetric cuts on the transverse momenta of the leptons,
\begin{eqnarray}
q_T^\ell > \SI{25}{GeV} \,, \qquad \left| \eta_\ell \right| < 2.4\,, \qquad
 \SI{71}{\GeV} < m_Z < \SI{111}{\GeV}\,.
\end{eqnarray}
We begin by examining the situation at \NLO{} for the total cross-section.
We are first interested in the asymptotic behavior of the residual subleading-power terms defined 
in \eq{residual}, in particular to demonstrate the scaling indicated by the power $p$ discussed in 
\cref{sec:fiducialpc}.  
\begin{figure}
	\centering
	\includegraphics[width=0.49\columnwidth]{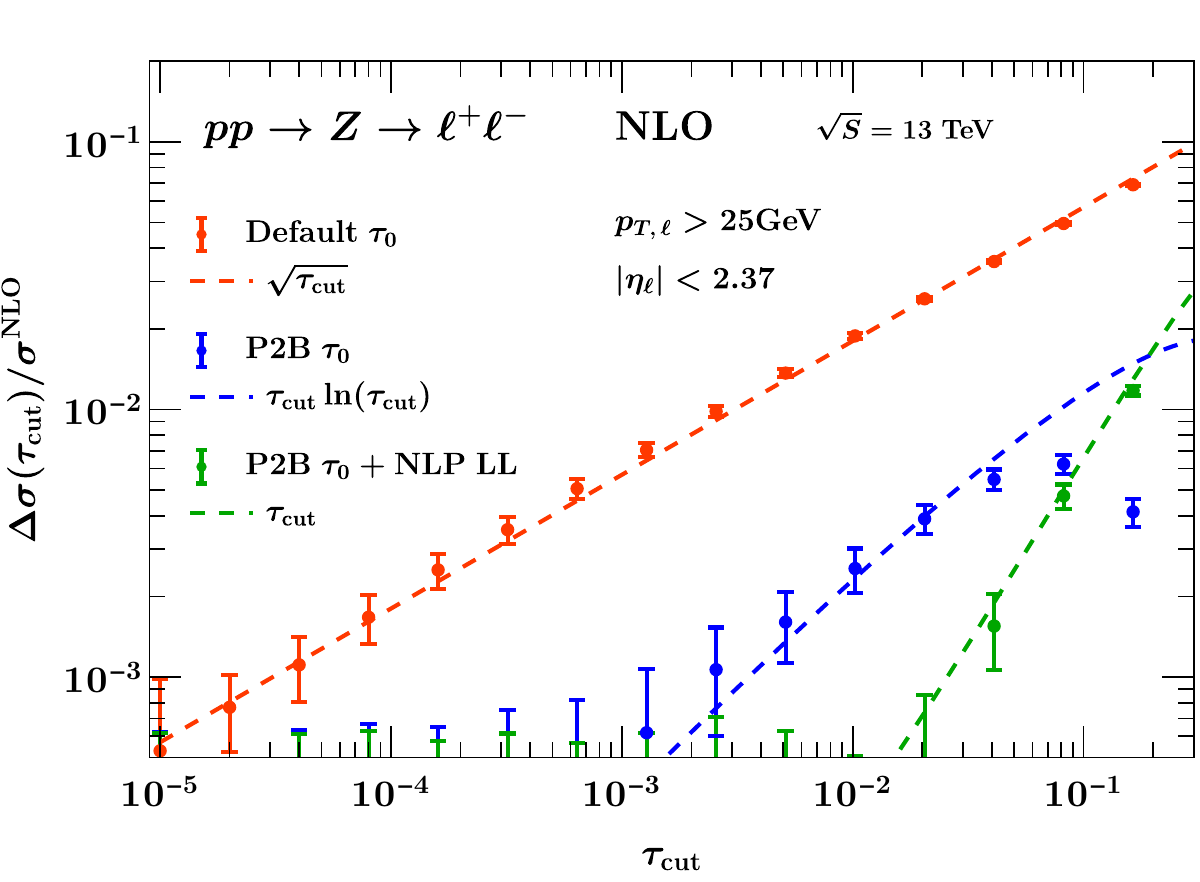}
	\includegraphics[width=0.49\columnwidth]{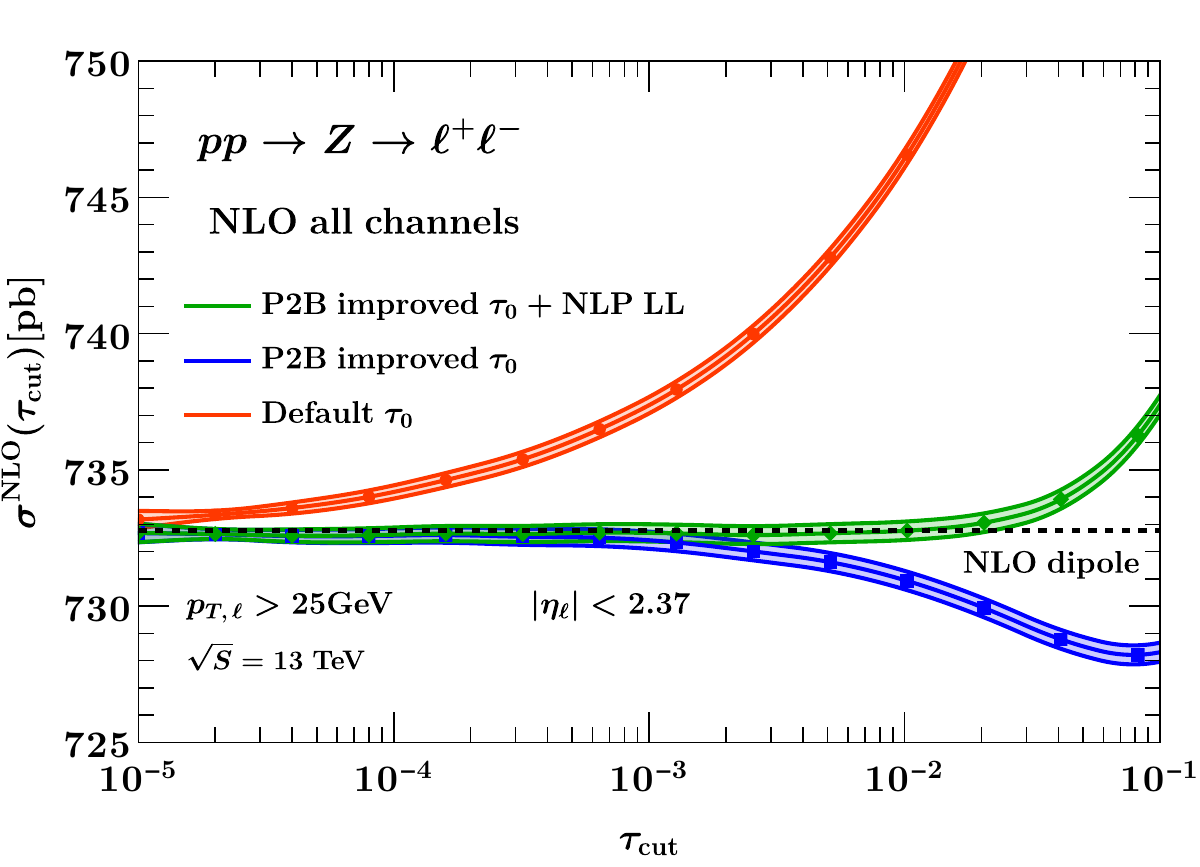}
	\caption{$Z \to \ell^+ \ell^-$ at \NLO{}: The left plot shows the slicing residual 
	$\Delta\sigma$ for $0$-jettiness subtractions and improvements thereon. The right plot 
	shows the total 
	cross-sections and a comparison to the local subtraction result. Interpolation and scaling 
	lines are shown 
	to guide the eye.}
	\label{fig:Z_NLO}
\end{figure}
We compute the slicing residual from the difference of the \NLO{} cross-section computed using 
local dipole subtraction and the one calculated in $0-$jettiness subtractions
at a given value of the $\tauc$ parameter.

Our numerical findings are shown in \fig{Z_NLO} (left). We see that the default
procedure suffers from fiducial $\sqrt{\tauc}$ power corrections, i.e. $p=1/2$, but that after
using the \PTB{} prescription these are removed and linear $\tauc$ 
scaling of the slicing residual is restored. This is in agreement with the findings of \refcite{Ebert:2019zkb}.
Once the fiducial power corrections are removed with \PTB{}, the 
next-to-leading-power leading-logarithmic ({\abbrev NLP LL}) corrections that are computed 
without fiducial cuts can be properly 
included, see \cref{sec:fiducialpc}. They lead to a substantial further decrease in the size of the 
residual.
The impact of the slicing residual on the calculation of the total \NLO{} cross-section in each of 
these
scenarios is summarized in \fig{Z_NLO} (right). 

\begin{figure}
	\centering
	\includegraphics[width=0.49\textwidth]{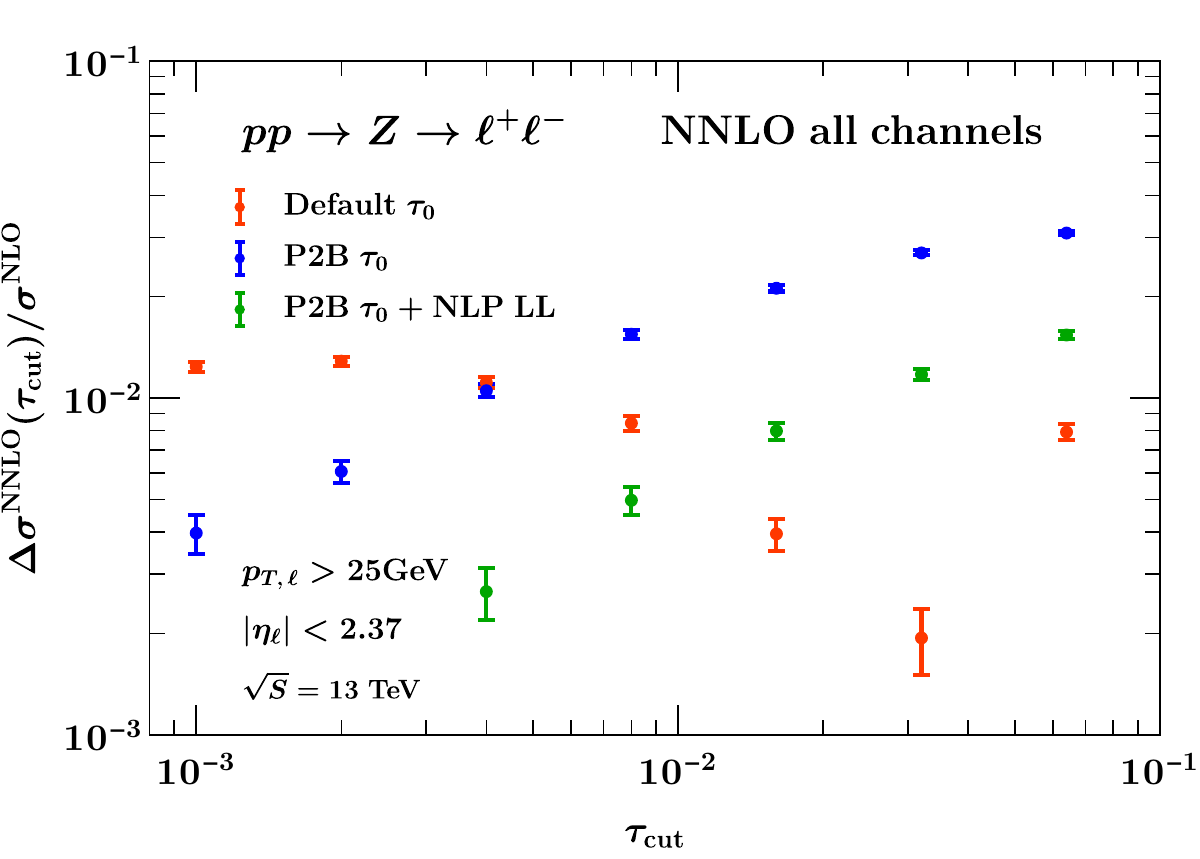}
	\includegraphics[width=0.49\columnwidth]{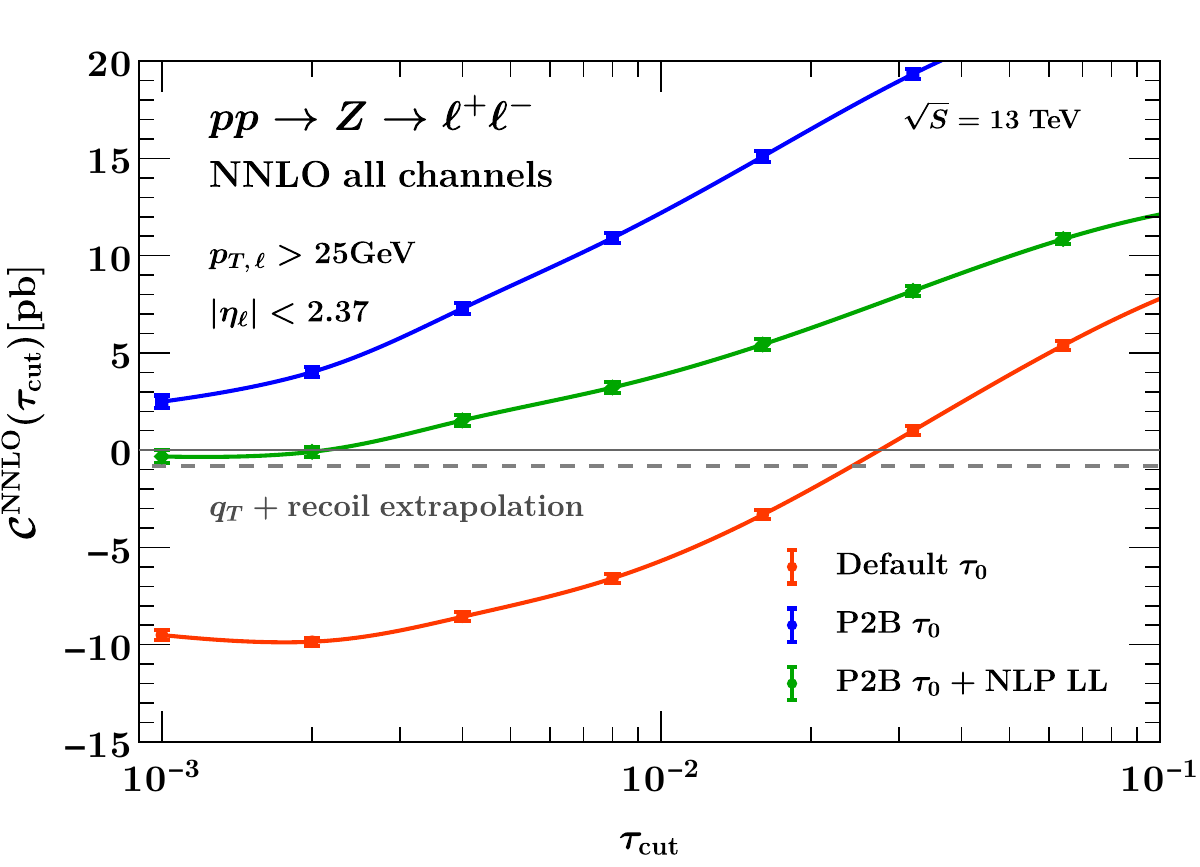}
	\caption{$Z \to \ell^+ \ell^-$ at \NNLO{}: 	
	The left plot shows $\Delta\sigma$ for various subtraction methods and improvements thereon. 
	The right plot shows the total cross-section coefficients ($\mathcal{C}_\NNLO{}$)  and a 
	comparison to the result obtained with $q_T$ subtractions with recoil.}
	\label{fig:Z_NNLO}
\end{figure}

The situation at \NNLO{} is summarized in \fig{Z_NNLO}. 
While no convergence is apparent at all for the default $0$-jettiness slicing, which suffers
from sizable $\sqrt{\tauc}$ power corrections, the \PTB{} corrections allow for cross-sections with per-mille level cutoff truncation errors. 
Our reference result is computed using $q_T$-subtractions with recoil  power corrections and is shown
as the dashed line. 
This result is obtained 
from a calculation using $q_{T\text{cut}}/Q=0.003$
that is in full agreement channel-by-channel with the corresponding result from
\MATRIX{} \cite{Grazzini:2017mhc,Buonocore:2021tke},
as reported in  \tab{MATRIXcomparison}.
\begin{table}
\centering
\begin{tabular}{|c|l|l|l|}
\hline
$pp \to Z \to \ell^+ \ell^- $ \NNLO{} coefficient  & \multicolumn{1}{c|}{$q\bar{q}+qq^\prime$} & 
\multicolumn{1}{c|}{$qg$} & \multicolumn{1}{c|}{$gg$} \\ \hline
\MCFM{} $q_T$ + recoil & \num{48732 \pm 316} fb & \num{-31819 \pm 175} fb & \num{13870 \pm 25} fb \\ \hline
\MATRIX{} $q_T$ + recoil & \num{48695 \pm 364} fb & \num{-31798 \pm 131} fb & \num{13786 \pm 205} 
fb \\ \hline
Relative Difference & \multicolumn{1}{c|}{\num{0.08 \pm 0.99} \%} & \multicolumn{1}{c|}{\num{0.07 \pm 0.69} \%} & \multicolumn{1}{c|}{\num{0.61 \pm 1.53} \%}  \\ \hline
\end{tabular}
\caption{Comparison of \NNLO{} corrections to the total fiducial $Z$ cross-section obtained via 
\MATRIX{} and \MCFM{} using $q_T$ subtractions with a recoil prescription. Results are broken down 
by different partonic channels.}
\label{tab:MATRIXcomparison}
\end{table} 

The inclusion of the {\abbrev NLP LL} corrections further improves the 0-jettiness result, leading 
to asymptotic flat results already at $\tauc=10^{-3}$.
For example, in practical applications one 
might aim for an accuracy of better than $0.5\%$ on the 
total \NNLO{} cross-section, corresponding to
$\sigma^{\NNLO{}} \times 0.005 = \SI{733}{pb} \times 0.005 \sim \SI{3.5}{pb}$. 
This cannot be obtained with the unimproved jettiness methods (nor $q_T$ subtractions) without a $\tauc$ value well below $10^{-4}$, which requires an extraordinary amount of computational effort and begins to enter the regime of numerical instability for double precision calculations.
However, when using \PTB{}-improved jettiness subtractions it can be reached using $\tauc \sim 2 
\times 10^{-3}$.
After including {\abbrev NLP LL} corrections, $\tauc \sim 10^{-2}$ suffices.
\begin{figure}
	\centering
	\includegraphics[width=.49\columnwidth]{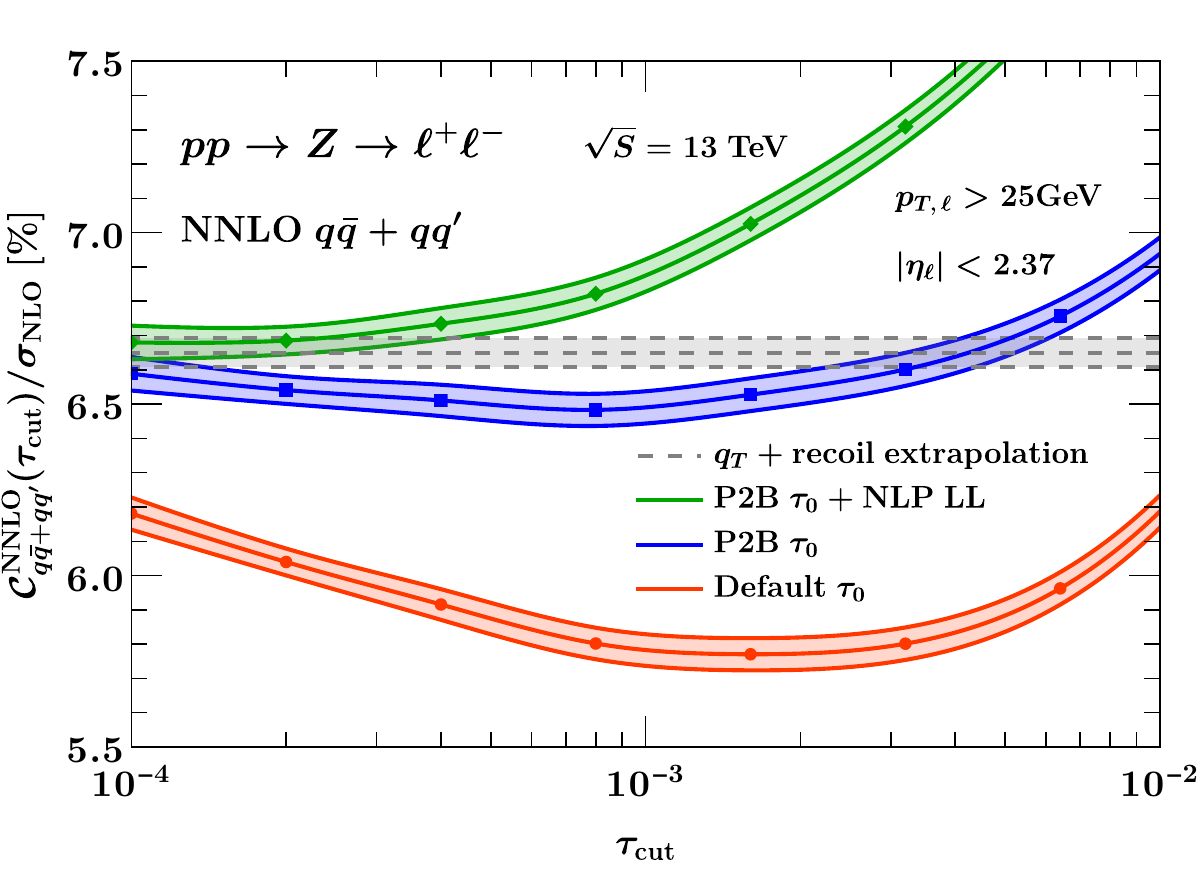}
	\includegraphics[width=.49\columnwidth]{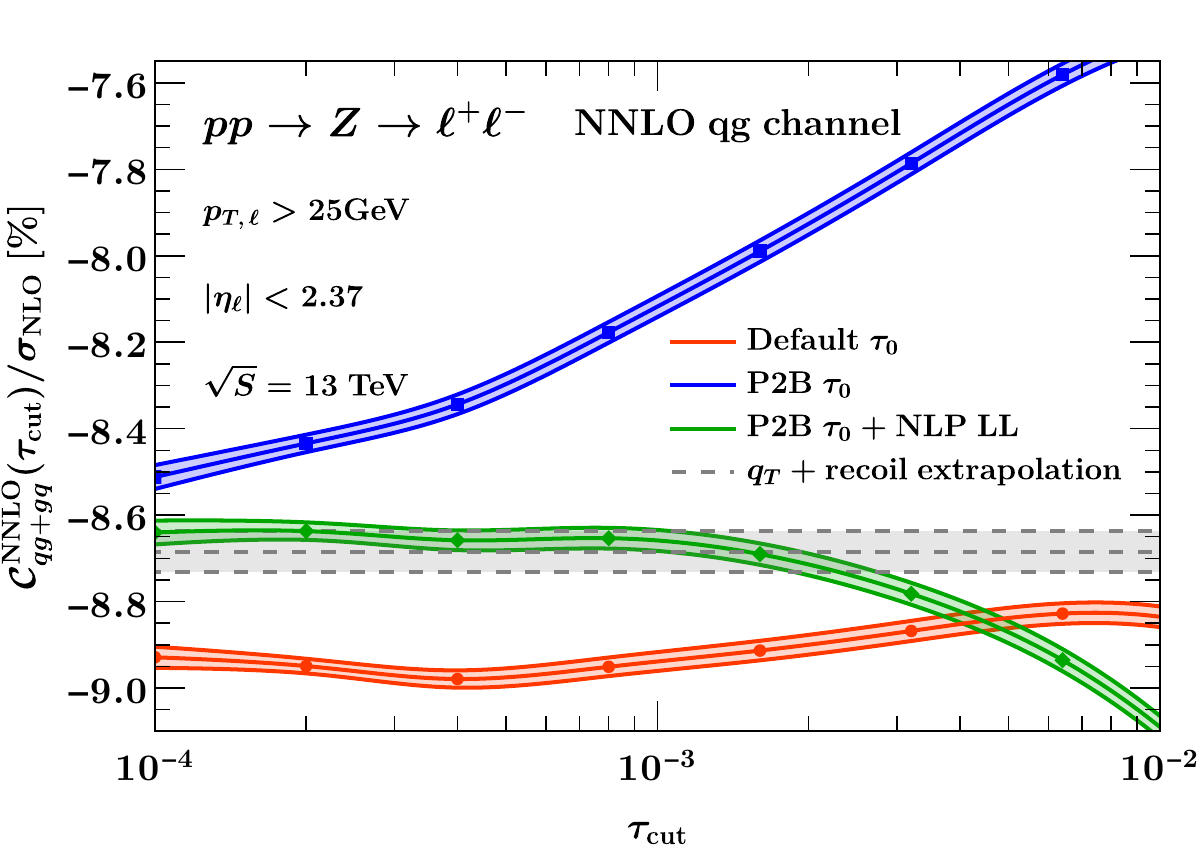}
	\caption{$Z \to \ell^+ \ell^-$: Dependence of the \NNLO{} total cross-section coefficient ($\mathcal{C}_\NNLO{}$) 
	on various subtraction improvements for the quark channel (left) and quark-gluon 
	channel (right). The gluon-gluon channel is smaller and its plot is left out here, but is 
	included 
	in the overall result in \cref{fig:Z_NNLO}. }
	\label{fig:Z_NNLO_coeff_channels}
\end{figure} 

In order to gain further insight into the behavior of the power corrections, we also analyze the
calculation of the \NNLO{} coefficient in two different combinations of channels: the $qg$ 
and the $q\bar q + qq^\prime$ channels.  This is especially instructive since the overall pattern
observed in \cref{fig:Z_NNLO} is obscured by the fact that the \NNLO{} coefficient is very small.
The breakdown into the two channels is shown in \fig{Z_NNLO_coeff_channels}. 
Again, we see that the removal of fiducial power corrections through the \PTB{} subtractions 
significantly reduces the dependence on $\tauc$.
We further observe that the {\abbrev NLP LL} terms dramatically improve the
convergence in the off-diagonal channel $qg$. This is the channel with the largest power 
corrections for this process and it therefore drives the improvement seen when including {\abbrev NLP LL}
corrections in \cref{fig:Z_NNLO}. 

For the diagonal channel the {\abbrev NLP LL} terms do not improve 
the size of the residual here, but they 
simply break the degeneracy that creates the erroneous plateau around $\tauc \sim 10^{-3}$. 
If, as in this case, the difference to the \PTB{}-improved result is large one could use this as a diagnostic to 
avoid being misled by a local extremum.
Overall, the size of the {\abbrev NLP LL} contributions can be used as 
another 
way to quantify the size of the slicing residual. 
The {\abbrev NLP LL} terms can also play an important role when asymptotically small values cannot 
be directly reached, but one instead has to rely on an extrapolation fit, for example in the case 
of limited computing resources or at higher orders. The additional simplification from removing the 
{\abbrev NLP LL} terms in the extrapolation formula can allow for improved asymptotic fits and 
a more precise extraction of subleading logarithmic terms.

\subsubsection{$Z$ rapidity distribution}
Having established the improvements of the \PtoBtau{}s on the total cross-section, 
we now look at the rapidity distribution of the dilepton pair as an example of a more differential distribution.
\begin{figure}
	\centering
	\includegraphics[width=0.49\columnwidth]{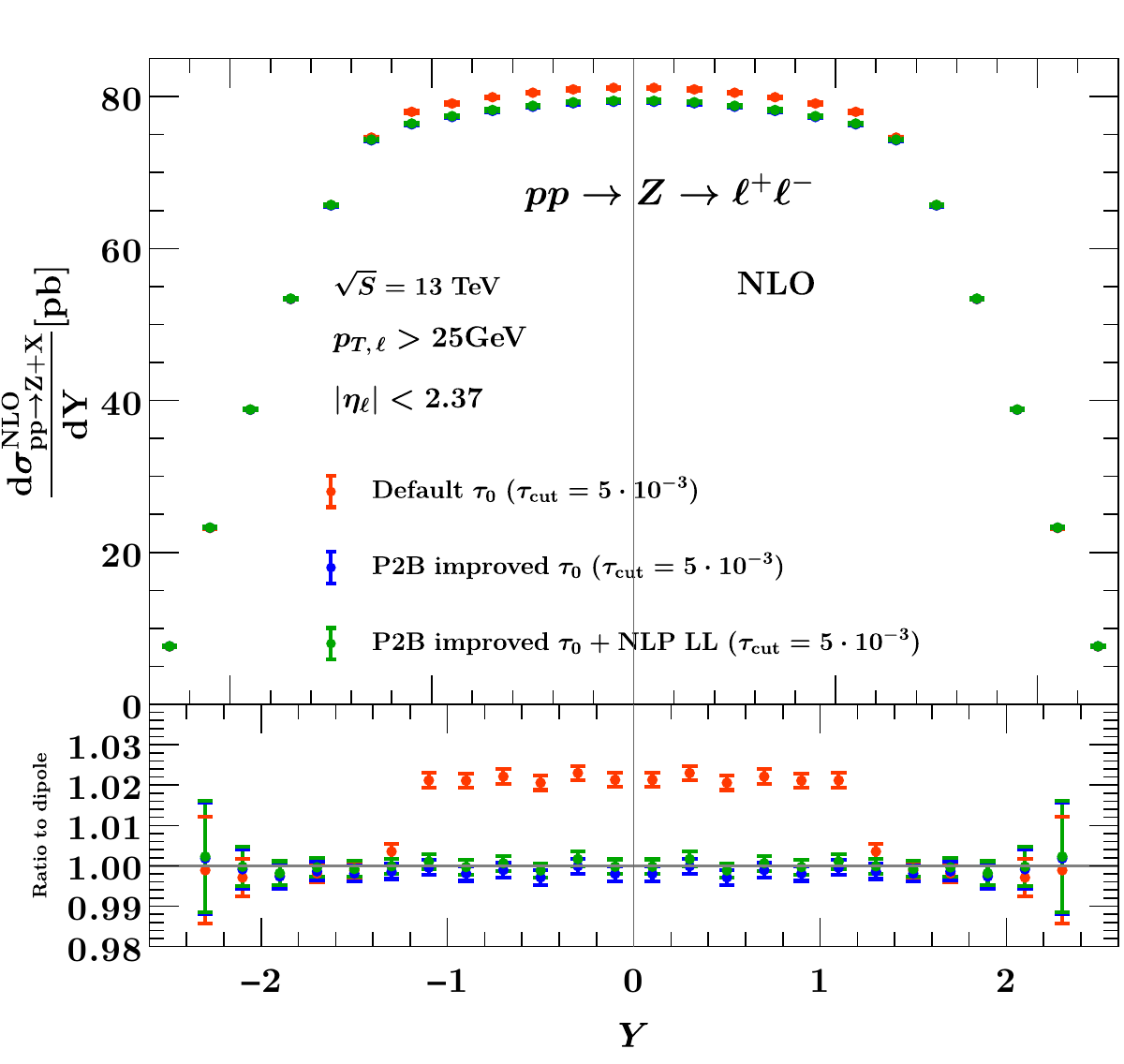}
	\includegraphics[width=0.49\columnwidth]{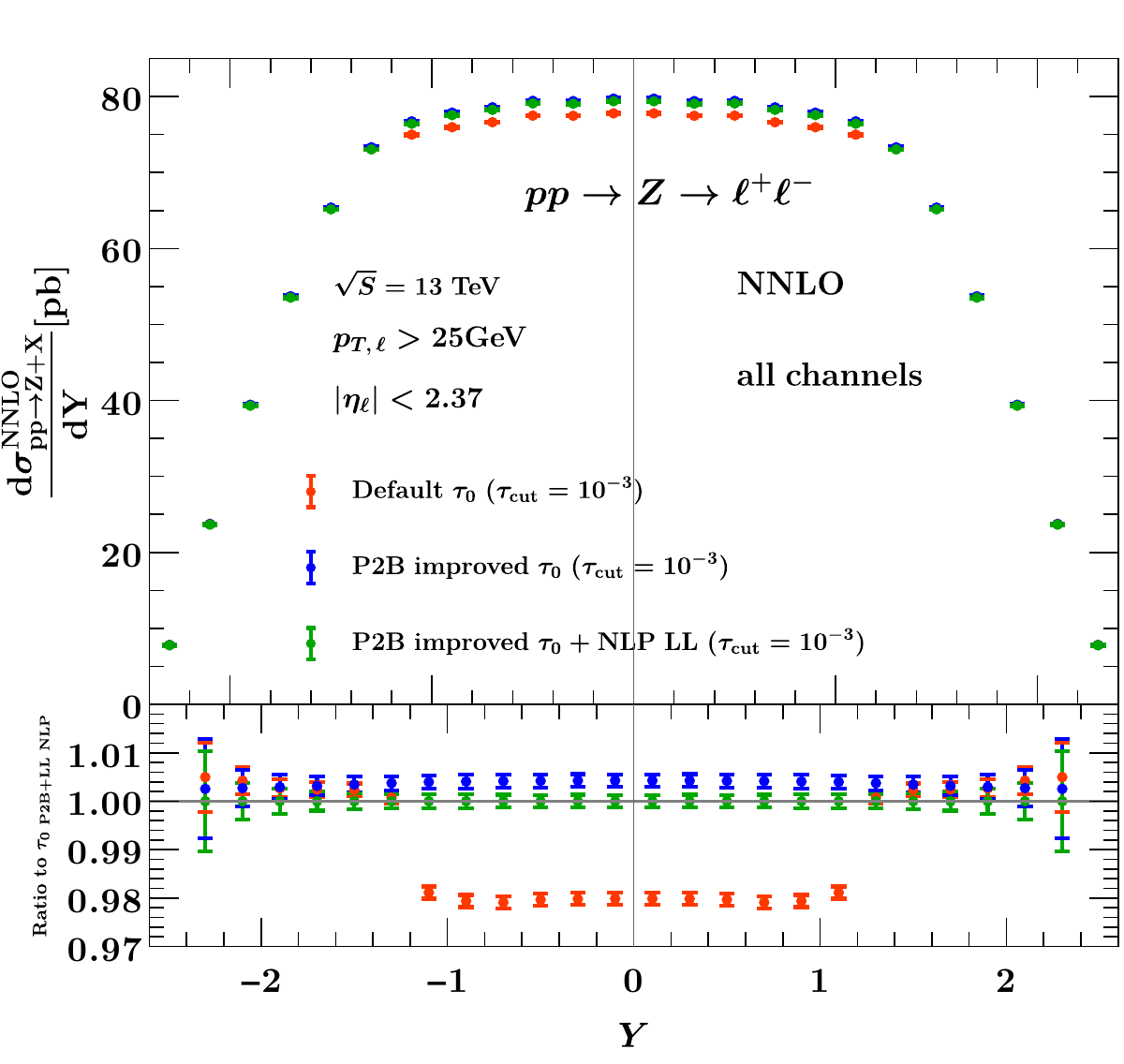}
	\caption{$Z \to \ell^+ \ell^-$: $Z$ rapidity distribution with symmetric lepton cuts at \NLO{} 
	(left) and \NNLO{} (right).}
	\label{fig:Z_NLO_rap}
\end{figure}

\begin{figure}
	\centering
	\includegraphics[width=1\columnwidth,trim={0 0 14cm 0},clip,trim={0 0 14cm 0},clip]{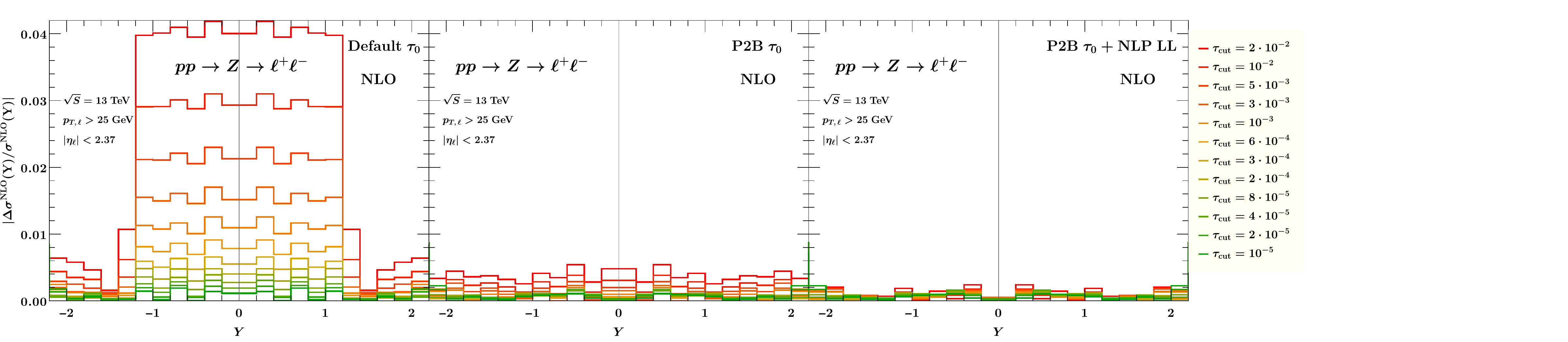}
	\caption{$Z \to \ell^+ \ell^-$: rapidity distribution residuals at \NLO{}. Each line 
	corresponds to a different value of the cut parameter and represents the residual error as a 
	function of Y, normalized to the \NLO{} distribution calculated using dipole subtraction.}
	\label{fig:Z_NLO_rap_residual}
\end{figure}

Results for the total $Z$ rapidity distribution at \NLO{} are shown in \fig{Z_NLO_rap} (left plot) 
for 
$\tauc=5\cdot10^{-3}$. We observe that the convergence with \PtoBtau{}s is excellent.
In \fig{Z_NLO_rap_residual} we show the slicing residuals at \NLO{} for the different methods.
We observe remarkable precision with the inclusion of \PTB{} corrections and further improvements 
when {\abbrev NLP LL} corrections are accounted for.

Results for the total $Z$ rapidity distribution at \NNLO{} are shown in \fig{Z_NLO_rap} (right 
plot) for 
$\tauc=10^{-3}$.  We see that without the \PTB{} corrections the size of the slicing residual is
about 2\%, flat across the distribution.  Once these are included the remaining correction from
the inclusion of {\abbrev NLP LL} terms is less than half a percent, again flat in $Y$. 

\begin{figure}
	\centering
	\includegraphics[width=1\columnwidth,trim={0 0 14cm 0},clip]{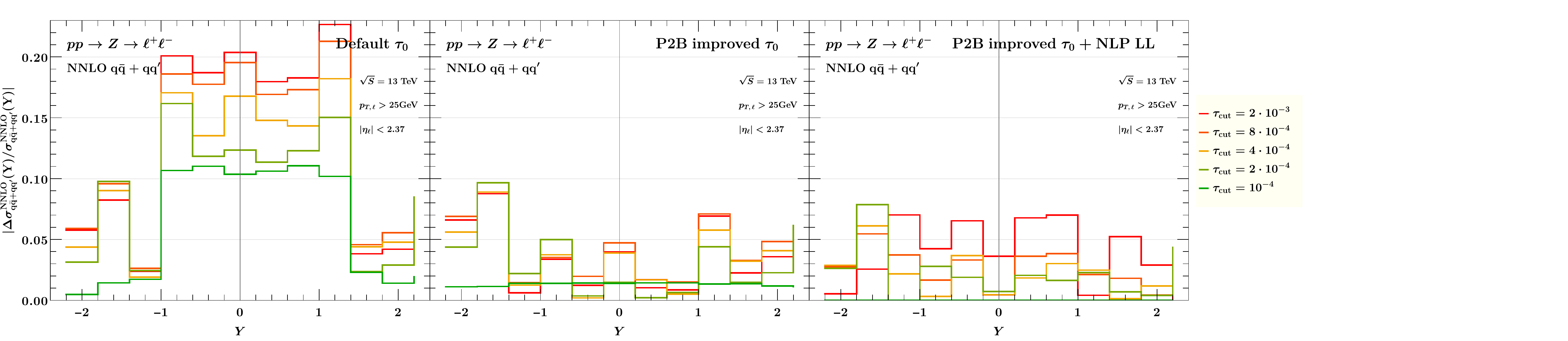}
	\includegraphics[width=1\columnwidth,trim={0 0 14cm 0},clip]{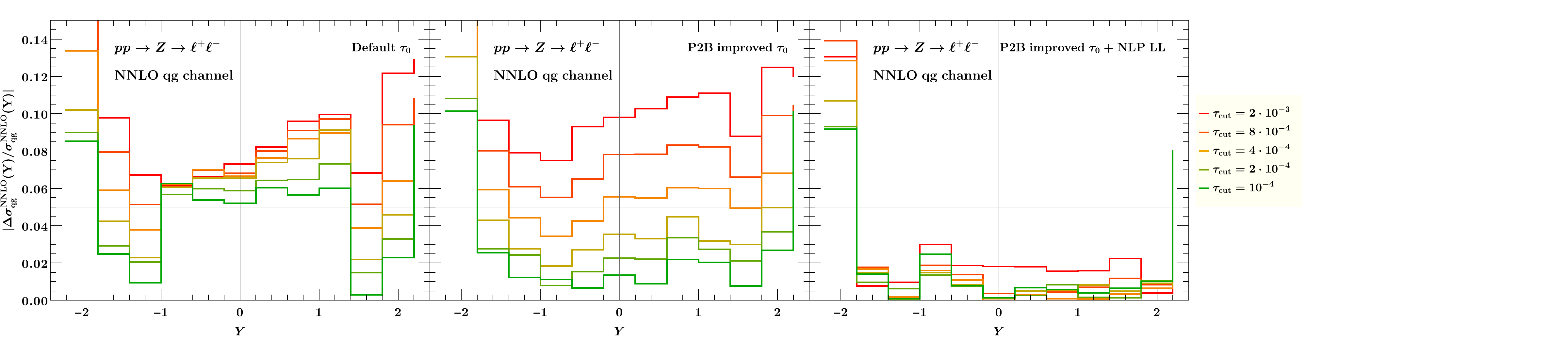}
	\caption{$Z \to \ell^+ \ell^-$: $Z$ rapidity distribution residuals at \NNLO{}, for the
	$q\bar q+qq^\prime$ channel (top) and $qg$ channel (bottom).}
	\label{fig:Z_NNLO_rap_channels}
\end{figure}

To better understand the $\tauc$-dependence of the total rapidity distribution we again present 
results in 
the
different channels at \NNLO{} in \fig{Z_NNLO_rap_channels}. 
We observe that the convergence of each separate channel is improved using \PtoBtau{} but that
the {\abbrev NLP LL} corrections are especially important for the $qg$ channel, as already observed
for the total cross-section. 
 Note that the $gg$ channel  does not receive any such 
corrections at this
order.

\subsection{Higgs production}
\label{sec:Higgs}

We consider Higgs production with decay into a photon pair, $H \to \gamma\gamma$, with a 
representative set of cuts:
\begin{eqnarray}
	&& p_{T,\gamma}^{\rm hard} > \SI{40}{GeV} \,,
	\qquad p_{T,\gamma}^{\rm soft} > \SI{30}{GeV} \,,
	 \qquad \left| \eta_\gamma \right| < 2.37 \,.
\end{eqnarray}

\begin{figure}
	\centering
	\includegraphics[width=0.48\columnwidth]{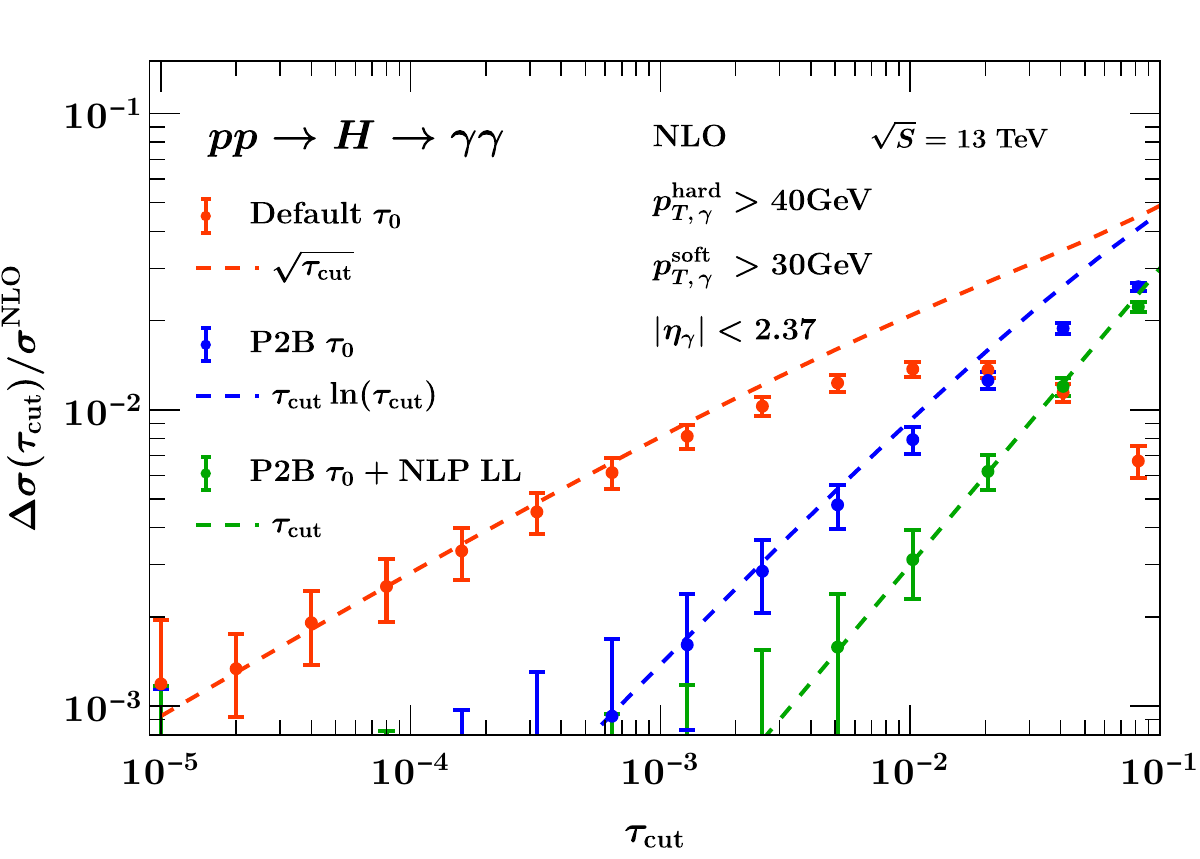}
	\includegraphics[width=0.48\columnwidth]{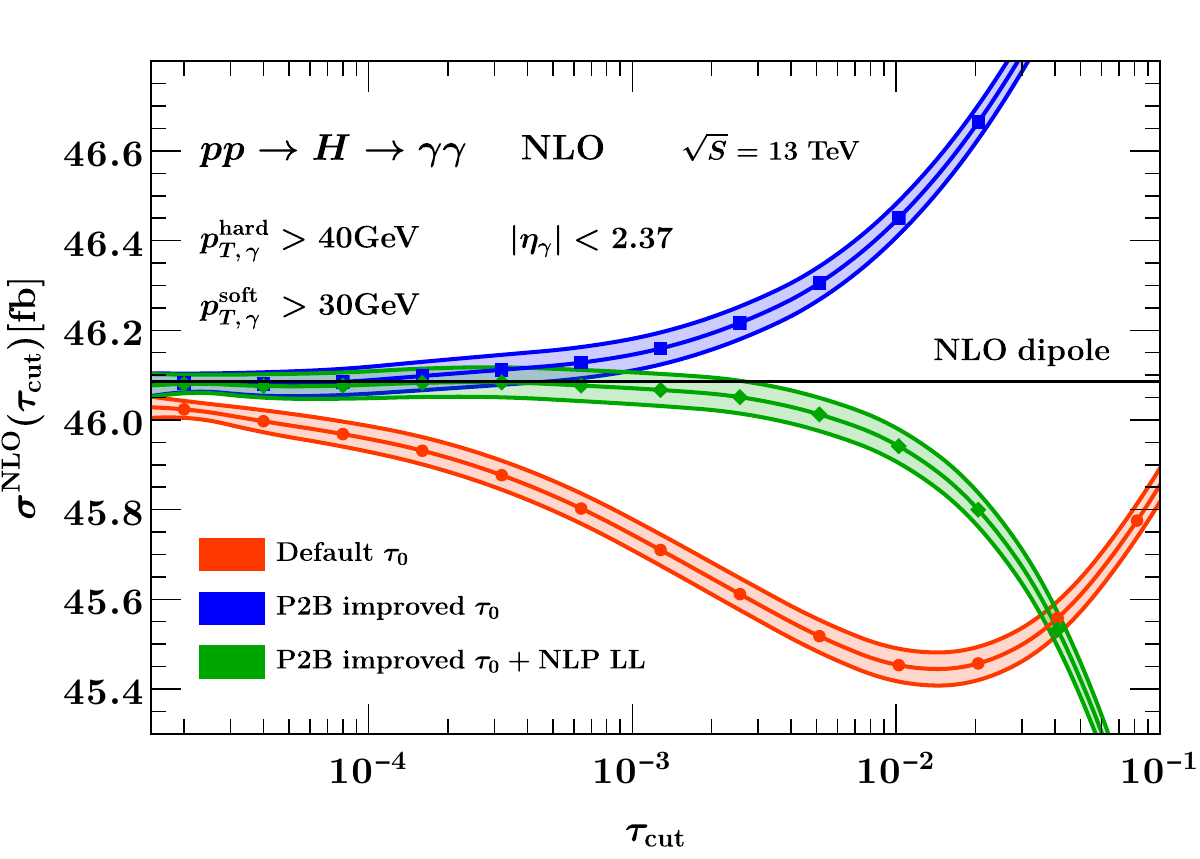}
	\caption{$H\to\gamma\gamma$ at \NLO{}: The left plot shows the slicing residual 	
	$\Delta\sigma$ for various subtraction method improvements. The right plot 
	shows the total 
	cross-sections and a comparison to the local subtraction result. Interpolation lines are shown 
	to guide the eye.}
	\label{fig:Higgs_NLO}
\end{figure}

The scaling behavior of the slicing residual at \NLO{} is shown in \cref{fig:Higgs_NLO}~(left). 
Including the \PTB{} corrections changes the scaling from $p=1/2$ to $p=1$, as expected.
Accounting for the {\abbrev{NLP LL}} power corrections decreases the overall size of the residual error.
The impact of these improvements on the \NLO{} cross-section calculation is demonstrated in \cref{fig:Higgs_NLO}~(right).
 For $0.1\%$ control of the \NLO{} result one only
needs $\tauc \sim 5 \times 10^{-3}$ (\PTB{}+{\abbrev NLP LL}) rather than $\tauc \sim 10^{-5}$
(unimproved).

\begin{figure}
	\centering
	\includegraphics[width=0.49\columnwidth]{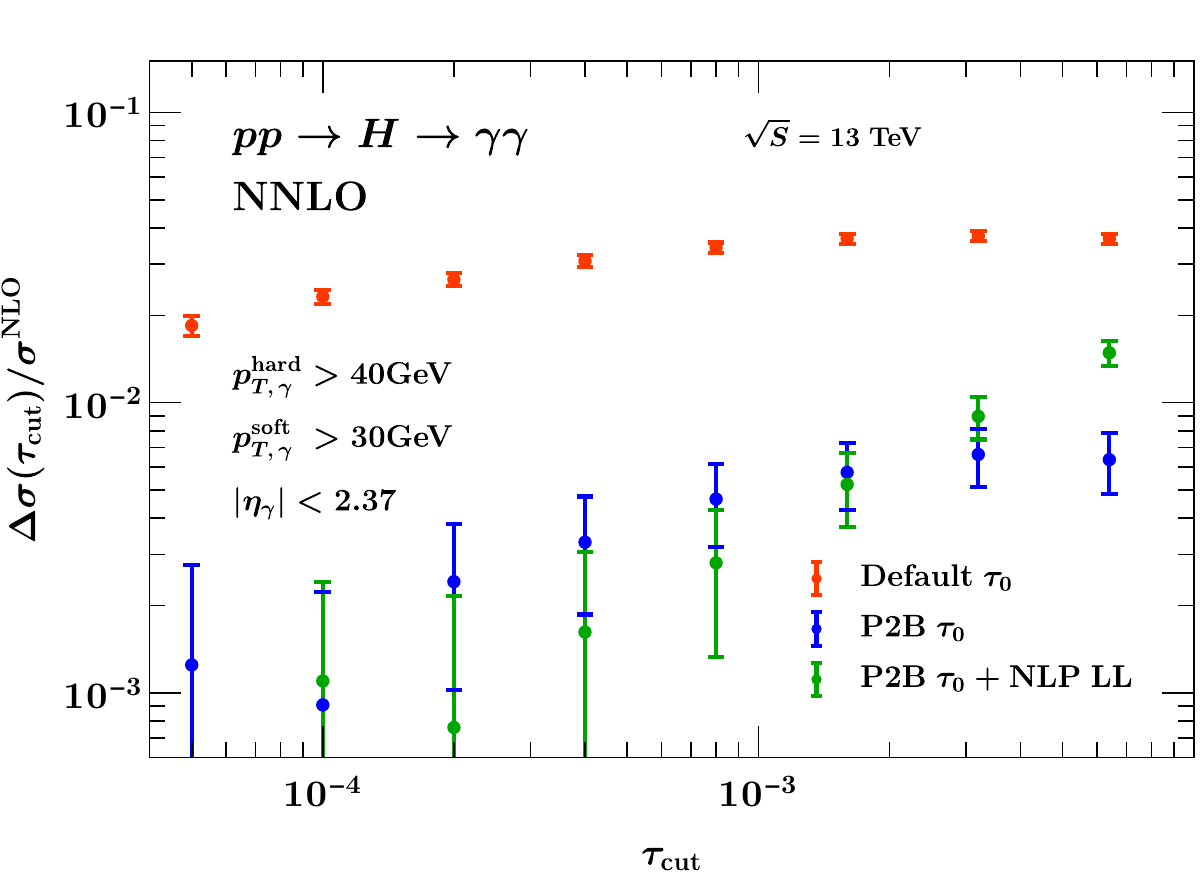}
	\includegraphics[width=0.49\columnwidth]{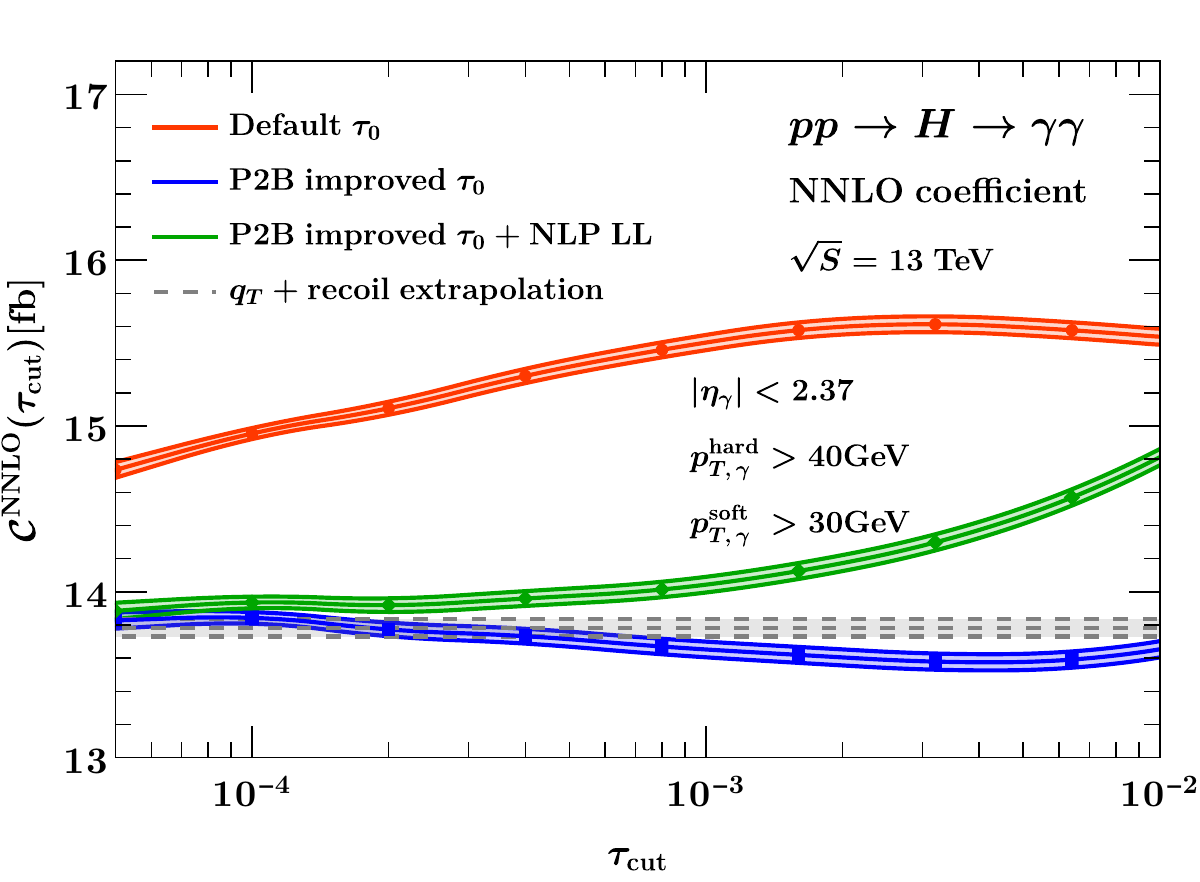}
	\caption{
	$H\to\gamma\gamma$ at \NNLO{}: The left plot shows the slicing residual $\Delta\sigma$ for 
	various subtraction method improvements. 
	The right plot shows the total cross-section coefficients ($\mathcal{C}_\NNLO{}$).}
	\label{fig:Higgs_NNLO}
\end{figure}

At \NNLO{} we obtain the results shown in \cref{fig:Higgs_NNLO}. It is clear that, without the inclusion
of any improvements, the default calculation has a large residual error and is only just beginning
to enter the region of asymptotic scaling. 
After inclusion of \PTB{} corrections the calculation is clearly in the asymptotic regime, with further
improvement in scaling after accounting for the {\abbrev NLP LL} corrections. 
These effects obviously result in significant improvements in the calculation of the \NNLO{} coefficient,
as shown in  \cref{fig:Higgs_NNLO}~(right).
In this case the
inclusion of {\abbrev NLP LL} corrections does not lead to substantial numerical improvements 
because the size of the power corrections accounted for by the \PTB{} scheme is already quite large.

\subsubsection{Higgs rapidity distribution}
\begin{figure}
	\centering
	\includegraphics[width=0.49\columnwidth]{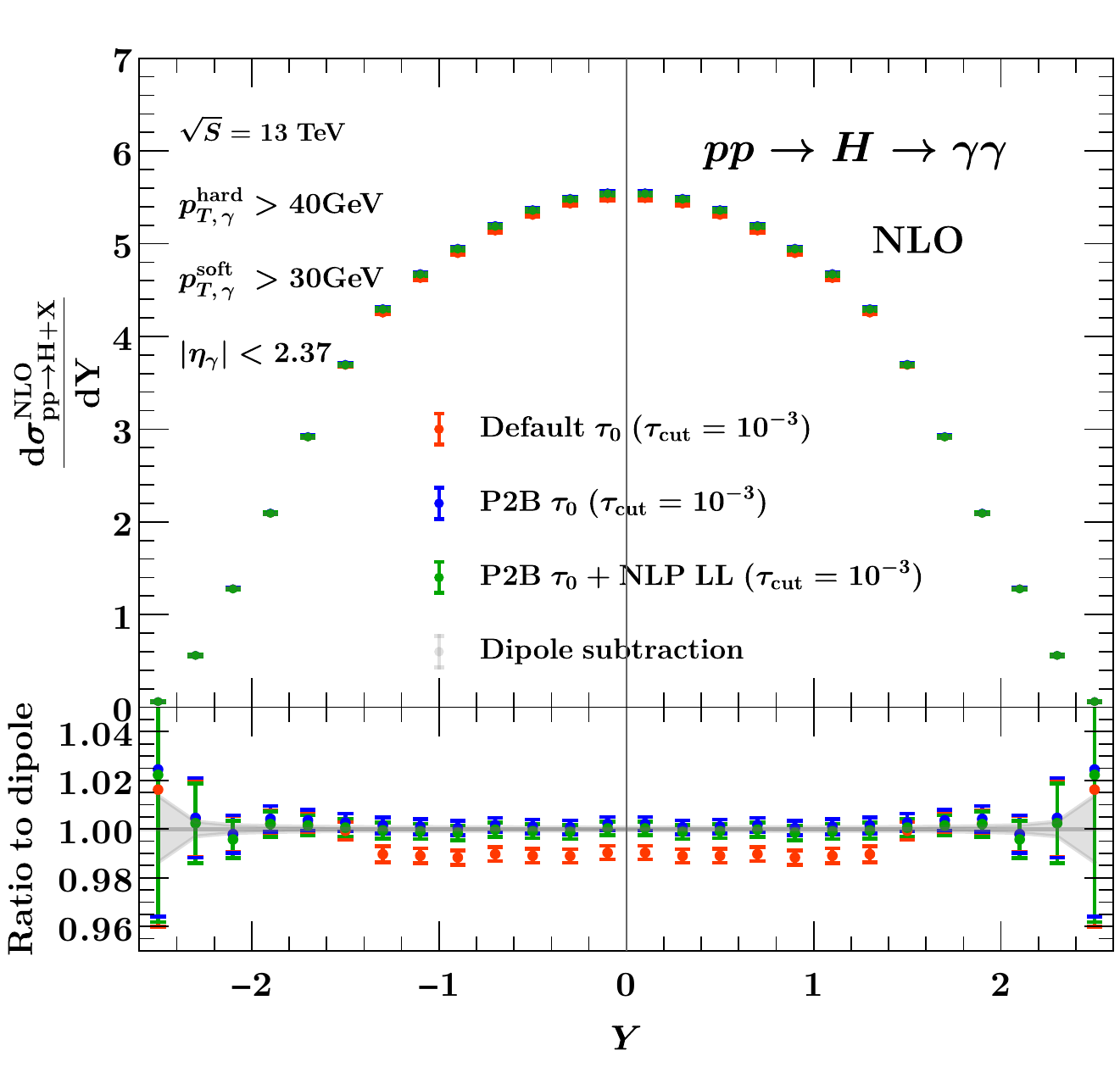}
	\includegraphics[width=0.49\columnwidth]{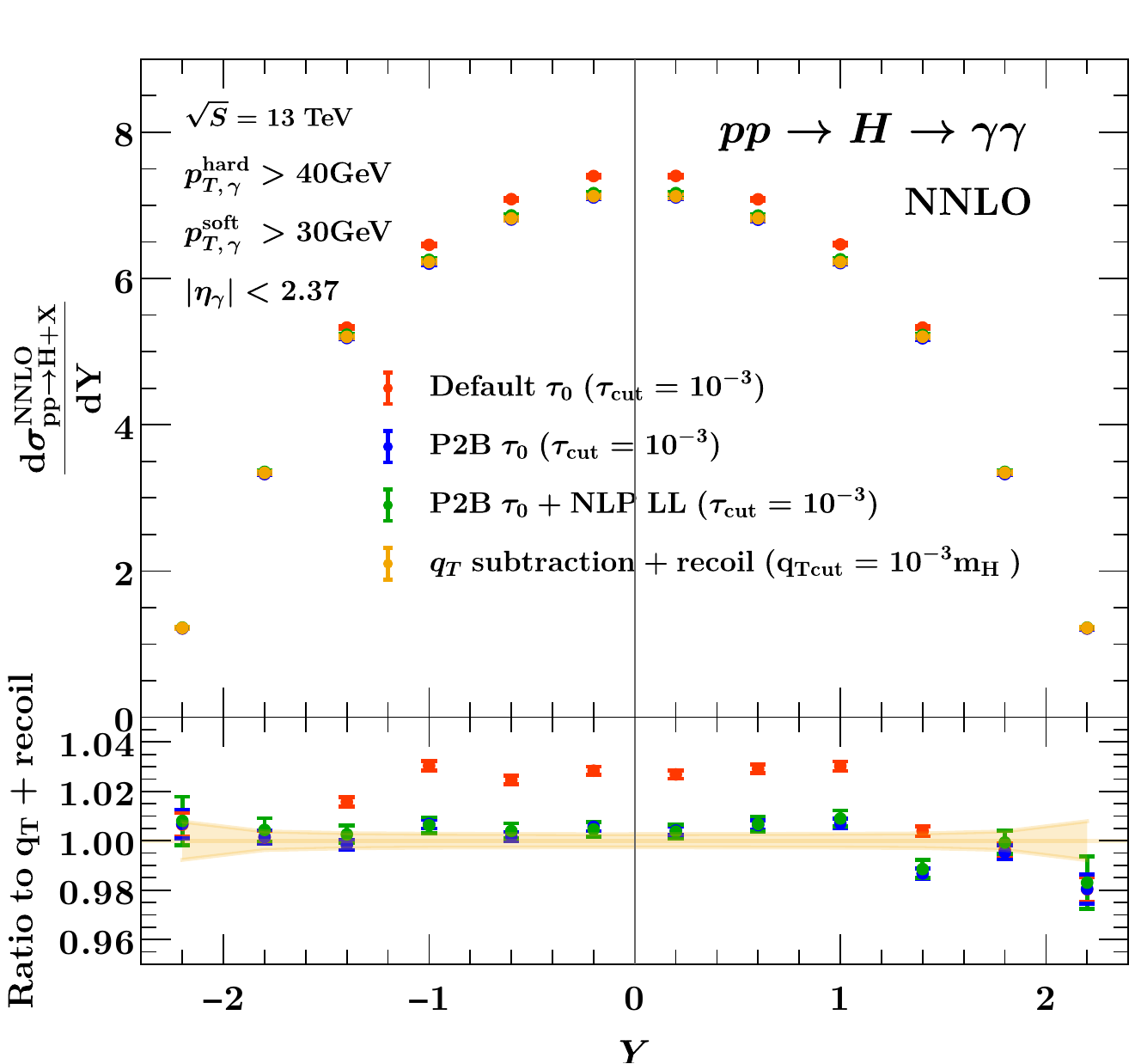}
	\caption{$H\to\gamma\gamma$: Higgs rapidity distribution at \NLO{} (left) and \NNLO{} (right),
	computed using $\tauc = 10^{-3}$ in the default, \PTB{} and \PTB{}+{\abbrev NLP LL} schemes.}
	\label{fig:Higgs_NNLO_rap_a}
\end{figure}
We now turn to a more differential quantity, the Higgs rapidity distribution. 
In \fig{Higgs_NNLO_rap_a} we show the total result at \NLO{} and \NNLO{} for $\tauc=10^{-3}$ in the different approaches.  
At \NNLO{} the unimproved result shows a flat slicing residual of about 3\%. 
Once \PTB{} corrections are included, the distribution agrees with our $q_T$+recoil reference result within numerical errors.
Improvements from including {\abbrev NLP LL} terms are small at the given $\tauc$.

\begin{figure}
	\centering
	\includegraphics[width=1\columnwidth,trim={0 0 14cm 0},clip]{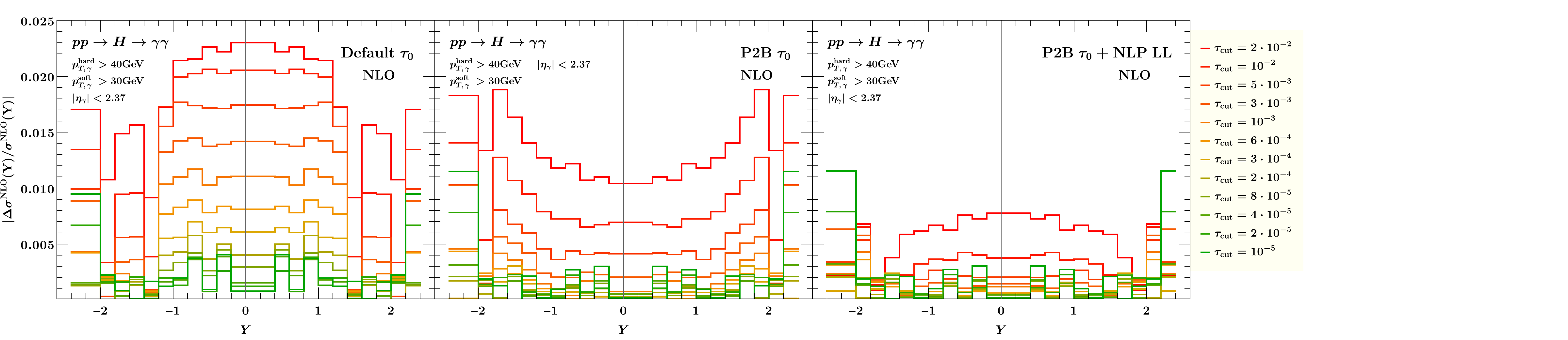}
	\caption{$H\to\gamma\gamma$: residual errors in the Higgs rapidity distribution at \NLO{}, 
	computed
	by comparing to a calculation performed using local dipole subtraction.}
	\label{fig:Higgs_NLO_rap}
\end{figure}

\begin{figure}
	\centering
	\includegraphics[width=1\columnwidth,trim={0 0 14cm 0},clip]{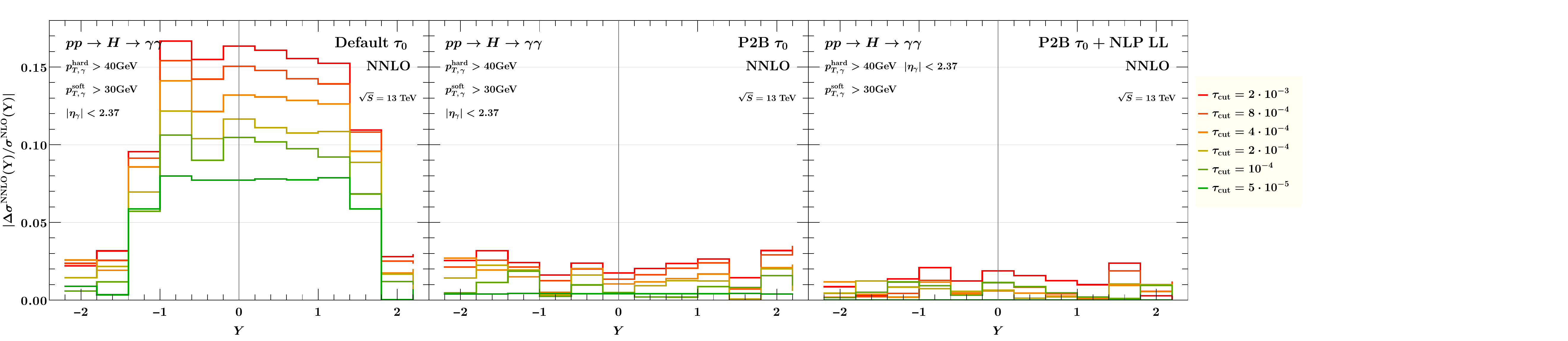}
	\caption{$H\to\gamma\gamma$: residual errors on the Higgs rapidity distribution at \NNLO{}.
	Each line corresponds to a different value of the cut parameter and represents the residual error as a 
	function of $y$, where
	the reference calculation is the one using \PTB{} and {\abbrev NLP LL} terms at 
	$\tauc=10^{-4}$.}
	\label{fig:Higgs_NNLO_rap}
\end{figure}

A similar picture emerges by looking at the slicing residuals at \NLO{} and at \NNLO{} for various 
values of the cut parameters, which are shown in \cref{fig:Higgs_NLO_rap} and 
\cref{fig:Higgs_NNLO_rap}, respectively.  
At \NLO{} the residuals are computed relative to the local dipole subtraction, 
while at \NNLO{} our reference is the calculation that employs \PTB{} and includes {\abbrev NLP LL} 
corrections with 
$\tauc=10^{-4}$.

As expected from the discussion of the total cross-section, there is significant improvement in 
convergence when using the \PTB{} corrections.
At \NLO{}, for $\tauc = 10^{-3}$ the residual error is approximately flat at 1\% in the unimproved 
case, a couple per mille once \PTB{} corrections are included, then completely negligible (within 
numerical uncertainties)  after including {\abbrev NLP LL} effects.
At \NNLO{} we see again that the  \PTB{} corrections provide a huge improvement in the convergence, 
with a more modest further gain after accounting for {\abbrev NLP LL} terms.

\section{Di-photon production at NNLO with $\PTB_\gamma$ $\tau_0$ and  $\PTB_\gamma ~q_T$ subtractions}
\label{sec:diphoton}
The \PTB{} method described in the previous sections directly captures fiducial power corrections for color-singlet processes.
However, for di-photon production with photon isolation, additional power corrections due to the 
isolation prescription contribute.
The goal of this section is to understand how to incorporate some of these isolation power 
corrections into the subtraction term for the non-local subtractions, improving their precision for 
photon processes. 

In \sec{diphotonpc}, we briefly review the behavior of power corrections for photon processes. 
Then in \sec{p2bgamma}, we discuss their interplay with \PTB-improved subtractions and present 
$\PTB_\gamma$-improved non-local subtractions, a new method to simultaneously incorporate kinematic 
power corrections and a set of isolation power corrections in the $\tau_0$ and $q_T$ subtractions.

\subsection{Power corrections in the presence of photon isolation}
\label{sec:diphotonpc}
In the case where photon isolation is included, the scaling of the power corrections depends on the 
isolation criterion in a non-trivial way. 
We consider Frixione's isolation \cite{Frixione:1998jh} for prompt photon production, with cone 
size $R$, exponent $n$, and transverse energy threshold $E_T^\iso$, which imposes the following 
requirement on the final state partons 
\be\label{eq:Frixionecut}
	\sum_{d(i,\gamma) \leq r} E_T^i \leq E_T^\iso \left[\frac{1-\cos(r)}{1-\cos(R)}\right]^n 
	\,,\qquad \forall r \leq R\,.
\ee
Sometimes the isolation energy is taken as a fraction of the photon transverse momentum, in which 
case $E_T^\iso = \varepsilon^\iso p_T^\gamma$. The angular distance measure $d(i,\gamma)$ between 
photon and parton $i$ is defined via $d(i,\gamma)=\sqrt{(\eta_i - \eta_\gamma)^2 + (\phi_i - 
\phi_\gamma)^2}$ with rapididies $\eta$ and azimuthal angles $\phi$.

Here one has to distinguish between the case of a quark or a gluon in the isolation cone. 
The case of a gluon in the isolation cone has been extensively discussed at \NLO{} in ref.~\cite{Ebert:2019zkb}. 
In the case of a quark (the actual fragmentation case), this smooth isolation prescription ensures the \QCD{} infrared-safe removal of the collinear quark-photon singularity.
The scaling of the power corrections in this case has been derived at \NLO{} in ref.~\cite{Becher:2020ugp}.
Generally one expects the scaling behavior to be similar at \NNLO{}, but this has not been studied in detail so far.
\begin{figure}
	\centering
	\includegraphics[width=0.47\columnwidth]{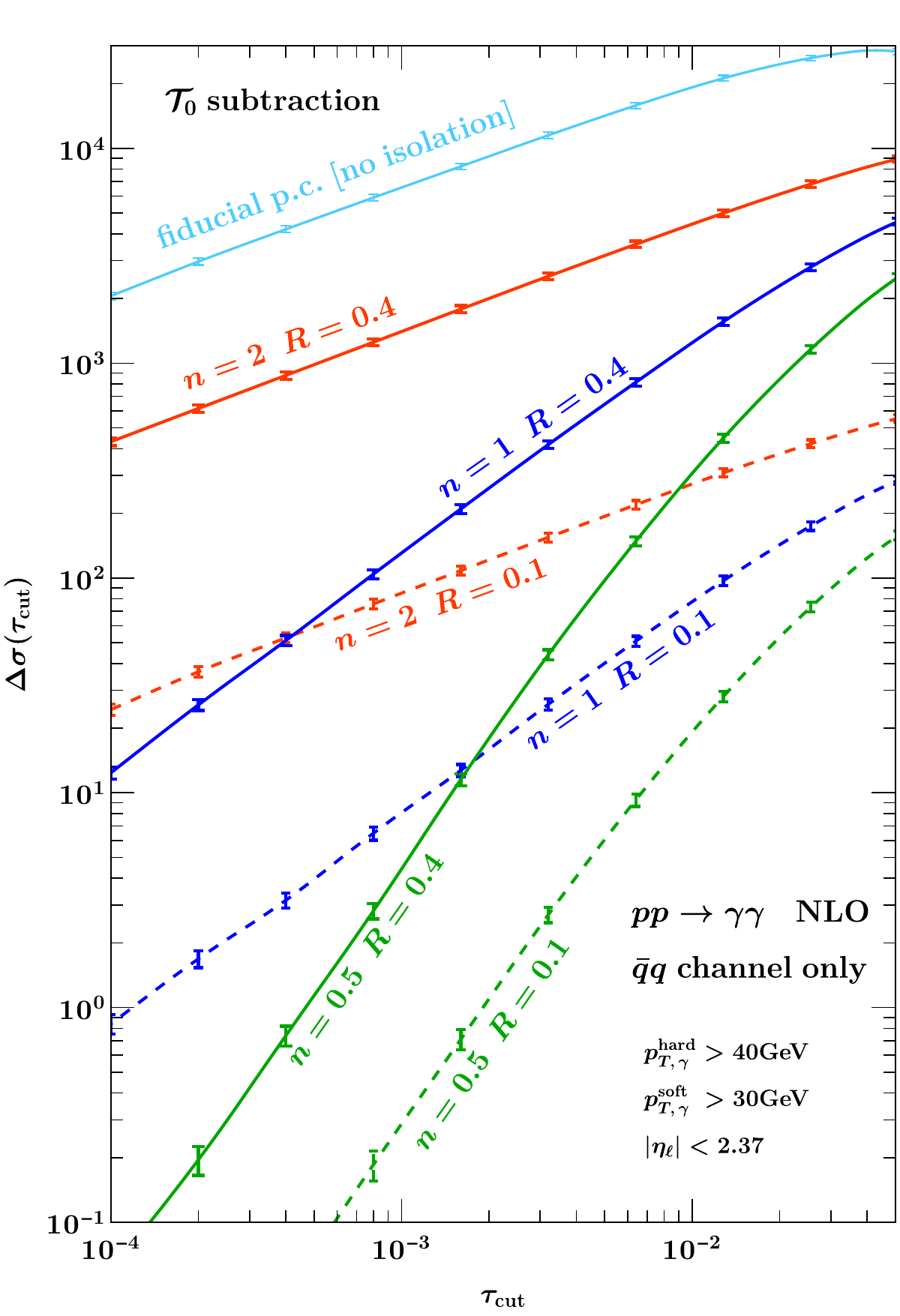}
	\includegraphics[width=0.485\columnwidth]{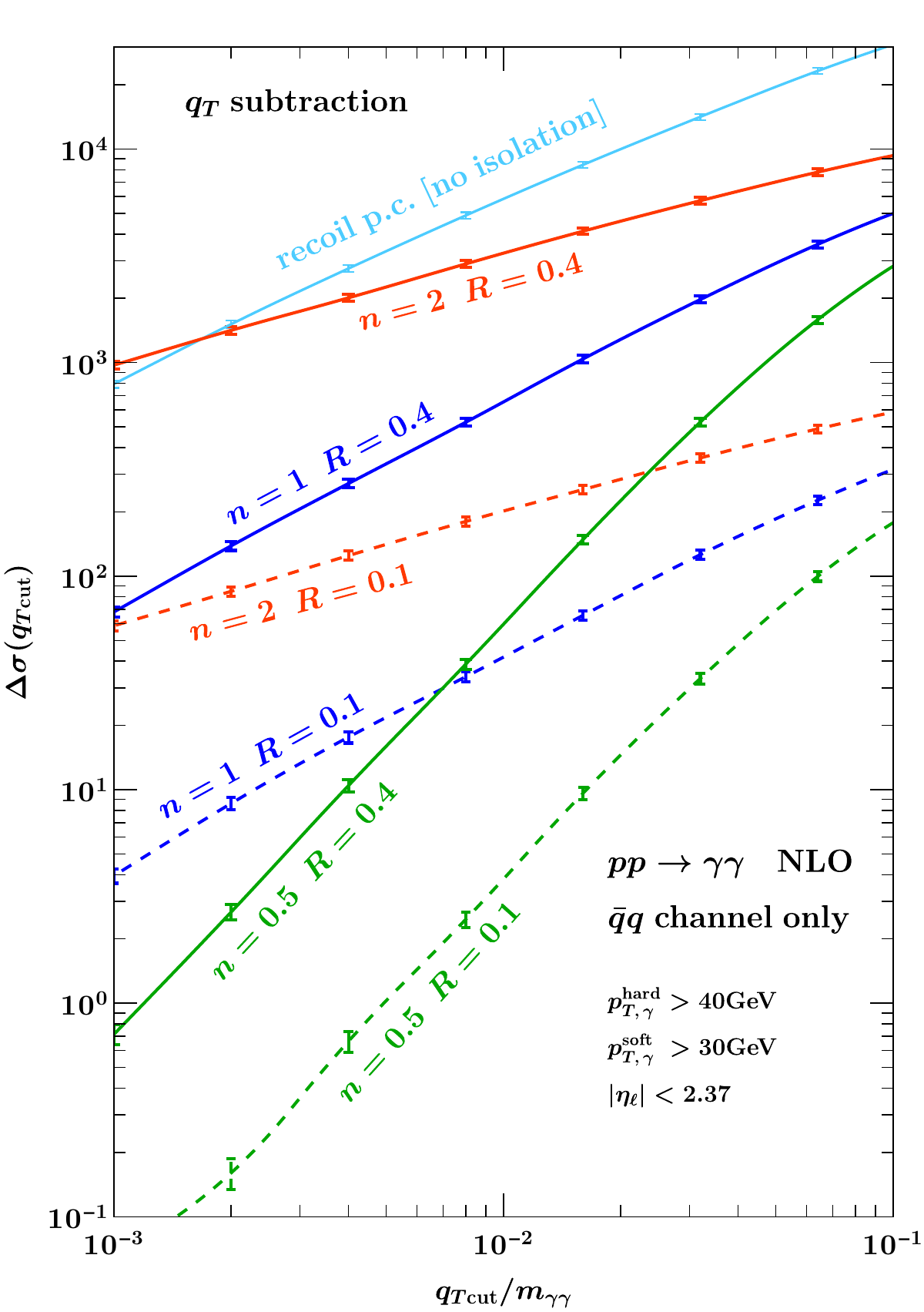}
	\caption{Difference between the \PTB{} correction with and without isolation for different isolation parameters at \NLO{} for the $q\bar{q}$ channel. Left: $0$-jettiness, 
	right: $q_T$. For comparison we also plot the \PTB{} correction without isolation.
	Note that the choice of colors in this figure is not related to the one in \eq{pccompletenequals1}.}
	\label{fig:diphoton_iso_scaling}
\end{figure}	
For a gluon, taking $R \ll 1$ and for $n\geq 1/2$, at \NLO{} one has 
\cite{Ebert:2019zkb,Becher:2020ugp}
\be\label{eq:pcFrixionecutgen}
	\text{gluon:}\qquad\Delta\sigma^{\mathrm{fiducial}}_{\mathrm{Frixione-iso}}(\rcut) \sim 
	\begin{cases} 
		R^2 \sqrt[n]{\Tauc /E_T^\iso} & \text{0-jettiness subtraction} \\
		R^2 \sqrt[n]{q_{T\mathrm{cut}}/E_T^\iso}& q_T\text{-subtraction}
	\end{cases}
\ee
or equivalently, 
\be\label{eq:pcFrixionercutgen}
	\text{gluon:}\qquad\Delta\sigma^{\mathrm{fiducial}}_{\mathrm{Frixione-iso}}(\rcut) \sim 
	\begin{cases} 
		R^2 \sqrt[n]{\rcut} \sqrt[n]{Q/E_T^\iso} & \text{0-jettiness subtraction} \\
		R^2 \sqrt[2n]{\rcut} \sqrt[n]{Q/E_T^\iso}& q_T\text{-subtraction}\,.
	\end{cases}
\ee
Here $Q$ is a hard scale of the process which is typically taken to be $m_{\gamma \gamma}$, the 
invariant mass of the di-photon pair.

To test \eq{pcFrixionercutgen}, we have calculated the difference of the \PTB{} correction for  $q\bar q \to \gamma\gamma$ cross-sections at 
\NLO{} with and without isolation for different isolation parameters using $\tau_0$ and $q_T$ subtraction as a function of the cut parameter. 
The results are presented in \fig{diphoton_iso_scaling} and show that the $n$-dependence follows \eq{pcFrixionercutgen}.
Moreover, as indicated, we see that changing the cone radius only impacts the overall size of the power correction but not its cutoff-scaling behavior. 

In the fragmentation case for a quark, at \NLO{} one has 
\be 
\text{quark:}\qquad\Delta\sigma^{\mathrm{fiducial}}_{\mathrm{Frixione-iso}} \sim
	\begin{cases} 
	         \frac{\Tcut}{Q} \log(R)  & \text{0-jettiness subtraction} \\
             \frac{q_{T\mathrm{cut}}}{Q} \log(R)& q_T\text{-subtraction}\,.
	\end{cases}
	\label{eq:pcFrixionequark}
\ee
The scaling in the $q_T$ subtraction case has been derived analytically in \refcite{Becher:2020ugp}
while we have determined the equivalent for jettiness subtractions numerically.

The net result is that in the sum of all partonic channels, where both gluon and quark final states
contribute, linear power corrections will always be present, irrespective of the value of $n$ that determines the scaling for the gluonic case. 
Typically, due to the scaling in $R$, the quark fragmentation power corrections dominate for small $R$, even when $n>1$.

\paragraph{Overall power corrections.}
Taking Frixione isolation with $n=1$ as an example, the sum of the hadronic, fiducial and photon 
isolation power corrections is as follows,
\be\label{eq:pccompletenequals1}
	\Delta\sigma^{\mathrm{fiducial}+\mathrm{hadronic}}_{p_T^\gamma+\mathrm{iso}, n=1} \sim 
	\begin{cases} 
		\kinematiccolor{\sqrt{\frac{\Tauc}{Q}}} +  \isocolor{\frac{\Tauc}{Q} \, 
		R^2\frac{Q}{E_T^\iso} + 
		\frac{\Tauc}{Q}\log(R)} + \inclcolor{\frac{\Tauc}{Q}}\,,   & \text{0-jettiness subtraction} 
		\\[.5cm]
		\kinematiccolor{\frac{q_{T\mathrm{cut}}}{Q}} + \isocolor{\frac{q_{T\mathrm{cut}}}{Q} \, R^2 
		\frac{Q}{E_T^\iso} + \frac{q_{T\mathrm{cut}}}{Q} \log(R)} + 
		\inclcolor{\frac{q^2_{T\mathrm{cut}}}{Q^2}}\,, & q_T\text{-subtraction}
	\end{cases}
\ee
where we have ignored the logarithmic $\tauc$ and $q_{T\text{cut}}$ behavior for brevity and have indicated the contributions coming from generic \kinematiccolor{$p_T^\gamma$ cuts}, \isocolor{photon isolation cuts}, and \inclcolor{hadronic/dynamical power corrections} with different colors. 
Note that power corrections for fiducial leptonic and photon isolation cuts have only been analyzed 
at \NLO{} \cite{Ebert:2019zkb,Becher:2020ugp}, although these corrections are expected to be 
dominant at higher orders as well. 
Further, as mentioned in \sec{fiducialpc}, the scaling of the hadronic power 
corrections (green in \eq{pccompletenequals1}) is based on a Drell-Yan 
analysis. A generalization has yet to be demonstrated in the literature, in particular for 
non-Born channels at higher orders.

Before moving on we now summarize the discussion of power corrections in \sec{fiducialpc} and here. 
We highlight the following:
\begin{itemize}
	\item In general, fiducial power corrections due to the kinematic cuts on the transverse 
	momentum of the final state leptons (or photons, e.g. for $H\to\gamma\gamma$ in gluon fusion), 
	scale as $\cO(\sqrt{\rcut})$, therefore dominating over the hadronic ones, which scale as $\cO(\rcut)$. 
	\item In the presence of photon isolation, power corrections are overall dominated by the 
	fragmentation contribution with a quark in the final state. 
	These further dominate over all other power corrections due to the $\log(R)$ scaling. \\
	For the gluonic channel there is a strong dependence of the power corrections on the isolation parameters and the subtraction method. 
	For example, with $n=1$ these isolation power corrections are important and behave as a leading source of power correction for $q_T$-subtraction, while for 0-jettiness they have a smaller impact, 
	comparable to that of hadronic power corrections.
	\item In scenarios where multiple scales are at play, power corrections to slicing schemes can acquire dependence on several of these scales. 
	Even for the simple case of color singlet production,  as one turns on fiducial cuts, the power 
	corrections may depend on scales such as $E_T^\iso$, $R$, etc. 
	While these scales formally enter as $\cO(1)$ parameters, they strongly change the numerical 
	impact of power correction compared to a naive analysis of linear or quadratic scaling. 
	It is therefore of paramount importance to be able to treat them systematically.
\end{itemize}

\subsection{Capturing fragmentation power corrections}
\label{sec:p2bgamma}
A crucial difference between di-photon production with photon isolation and Drell-Yan processes is 
that, when quarks are produced in the final state, the inclusive cross-section is ill-defined in 
perturbation theory in the absence of an isolation criterion or a non-perturbative object such as a 
fragmentation function.
For example, in $q_T$ subtractions with a recoil prescription for fiducial power corrections, it is 
clear that no photon isolation power corrections can be captured given that it only operates at the 
Born kinematics level.
While a naive application of the \PTB{} prescription is also not effective, as we describe 
below, the access to the full matrix element allows modifications of the scheme to capture a set of 
these power corrections.

To be more precise, in the case of the emission of a gluon into the isolation cone the \PTB{} 
prescription works as expected, as the inclusive cross-section is well defined even in the absence 
of an isolation criterion. 
With the isolation removing the gluon emission's phase space that is collinear to the photon, the soft-gluon singularity is canceled between $\mathcal{O}$ and $\mathcal{\tilde{O}}$ in the \PTB{} prescription, leaving behind a finite power correction that is captured.
In \cref{sec:diphotonnumeric} we demonstrate numerically that the \PTB{}-improvements in this case follow the expected power behavior and magnitude.
The power corrections from the \PTB{} terms match the power corrections obtained from a recoil prescription in $q_T$ subtractions.

On the other hand, in the case of the emission of a quark, the isolation prescription takes care of removing the collinear quark-photon singularity.
As the quark becomes soft the isolation ceases to act, but no soft \QCD{} singularity arises as the soft quark emission is power suppressed. 
A source of slicing residual is that in the slicing calculation such a contribution is neglected 
when the quark energy is smaller than the cut parameter.
Fortunately, we can correct for this effect in a simple way.
In realistic scenarios the photon has some transverse momentum cuts and, combined with the isolation, this implies that every quark that is close to the photon must be soft. 
Even below the cut, such quarks don't need any \PTB{} counterterm because their matrix element is not singular.
This implies that we can account for this effect by simply turning off the \PTB{} counterterm in the cone of radius $R$ around the photon. 
We demonstrate in \cref{sec:diphotonnumeric} that this procedure significantly improves the overall convergence, hence capturing a significant portion of the isolation power corrections.

Since this is a prescription for one emission, we apply it at \NLO{} and for the \NNLO{} real-virtual corrections with a quark emission. 
Further adjustments should be necessary to handle the double emission contributions and require 
further study. 
Nevertheless, we find that the prescription presented here significantly improves the cutoff dependence at \NNLO{} even if just applied to the real-virtual corrections, 
as presented in the results section. We refer to this improved scheme as $\PTB_\gamma$ subtractions, distinguishing it from the normal \PTB{} improved subtractions which do not include this prescription.

\subsection{Numerical results}
\label{sec:diphotonnumeric}

Following our discussion of photon-isolation power corrections, in this section we demonstrate the improvements
from the $\PTB_\gamma$-subtraction for di-photon production.
We use photon cuts as in $H\to\gamma\gamma$: $p_{T,\gamma}^{\rm hard} > \SI{40}{GeV}$, $p_{T,\gamma}^{\rm 
soft} > \SI{30}{GeV}$, $\left| \eta_\gamma \right| < 2.37$ and
Frixione photon isolation as in \cref{eq:Frixionecut} with  
$E_T^\text{iso}=\SI{10}{\GeV}$, $R=0.4$, and $n=1$. We show results for a 
center-of-mass energy $\sqrt{s}=\SI{13}{\GeV}$ with the \PDF{} set \texttt{NNPDF31\_nnlo\_as\_0118} 
\cite{NNPDF:2017mvq}. The factorization and renormalization scales are set to the invariant mass 
$Q$ of the di-photon system.
Because the distinction between final-state gluons (\PTB{}) and quarks ($\PTB{}_\gamma$) is crucial 
for the discussion of different schemes, we separate the $q\bar{q}+q q'$ and $qg$ (fragmentation) 
channels in our initial discussion.

Note that in all cases the smallest $q_{T,\text{cut}}$ and $\tauc$ are as low as can be achieved using numerical double precision, even using an improved treatment of all matrix elements in \MCFM{}~\cite{Bothmann:2024hgq}. 
Smaller values of $q_{T,\text{cut}}$ and $\tauc$ would require technical cutoffs that have an impact on the slicing cut itself. As will become clear, the inclusion of power corrections in the photon isolation itself is essential to achieve a reliable result.

\begin{figure}
	\centering
	\includegraphics[width=0.492\columnwidth]{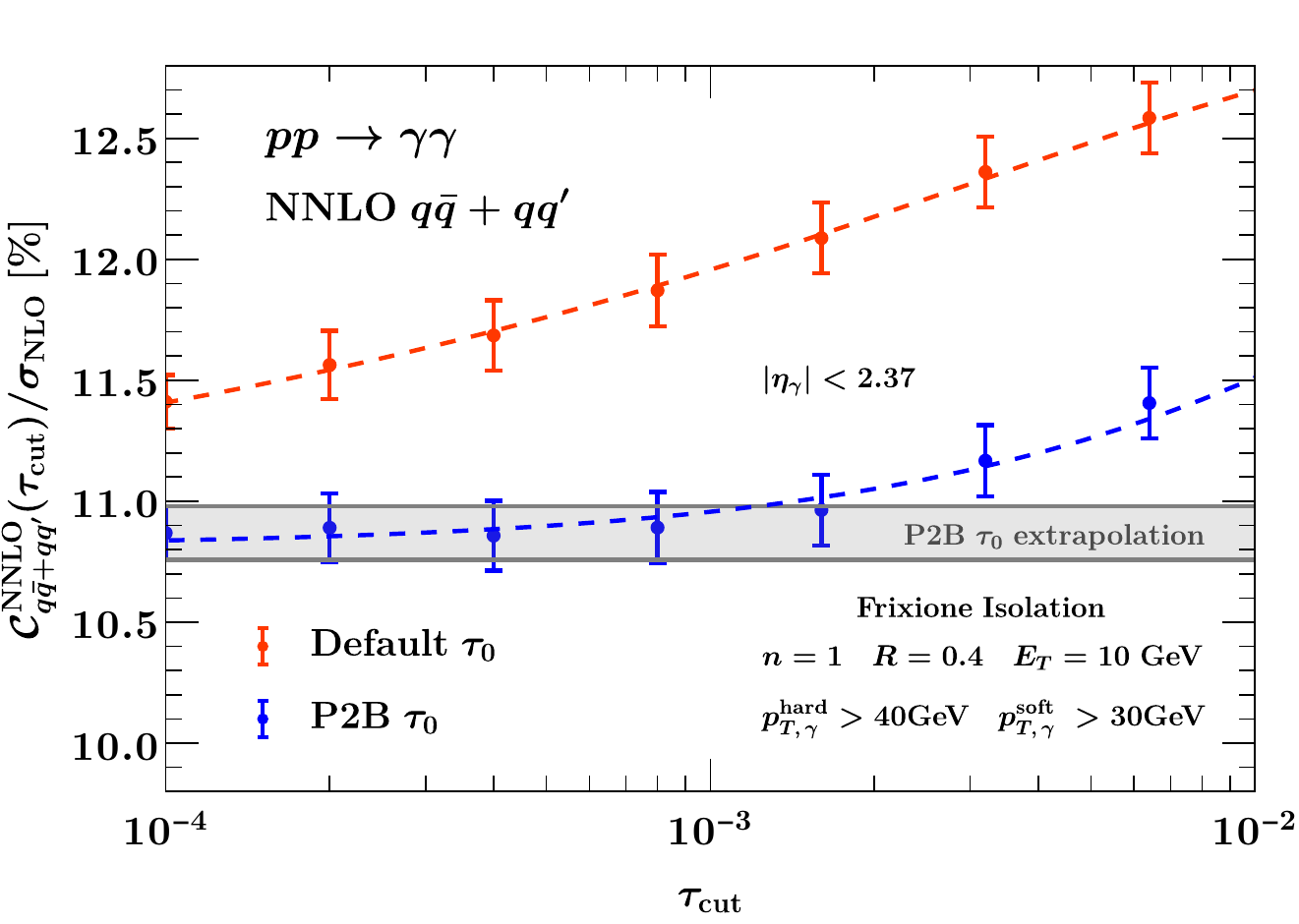}
	\includegraphics[width=0.492\columnwidth]{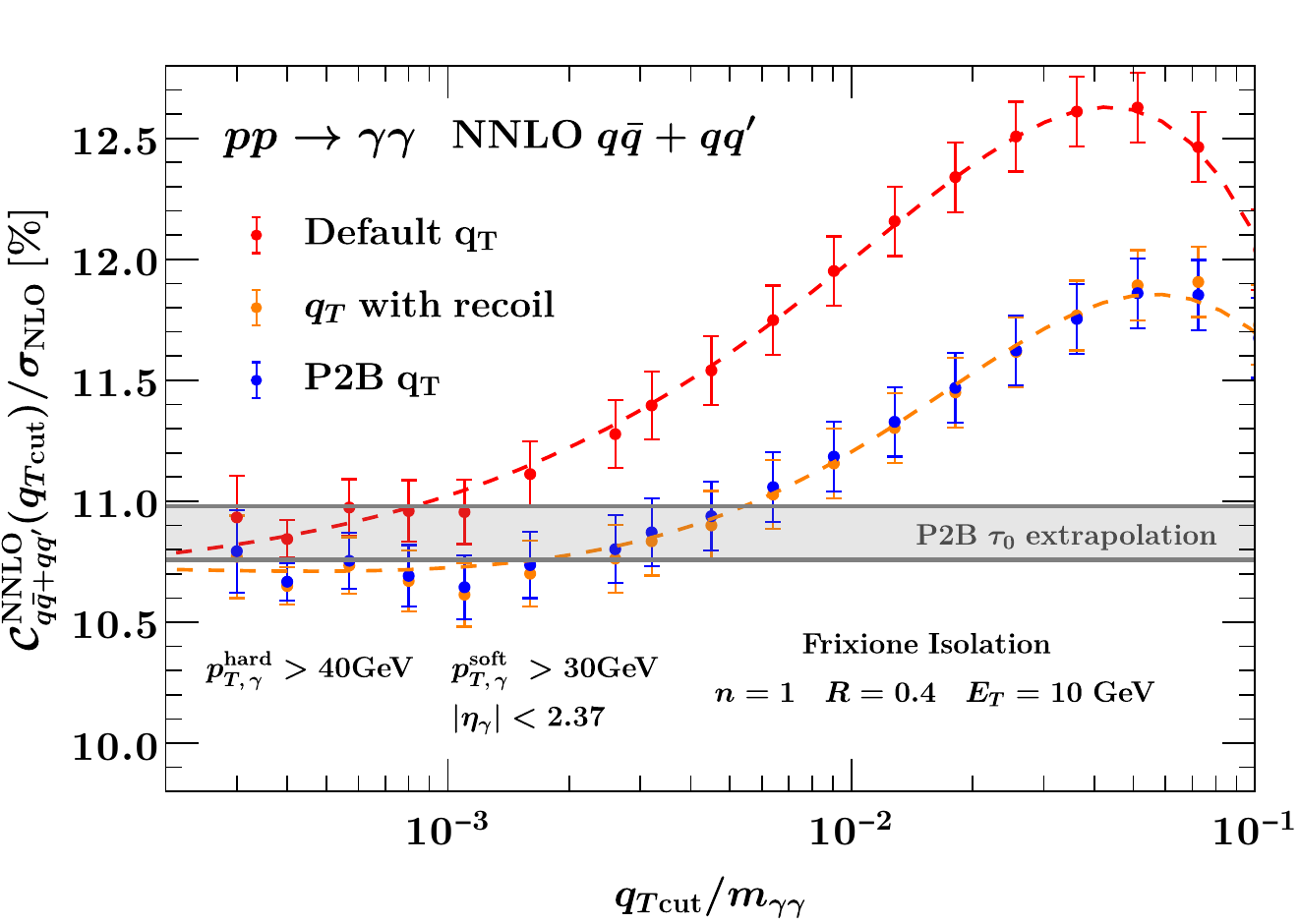}
	\caption{\NNLO{} coefficient K-factor for di-photon production, $q\bar{q}+q q^\prime$ channel 
	only, for 
	$0$-jettiness (left) and $q_T$ subtractions (right) and their improved versions.}
	\label{fig:diphoton_nnlo_qqbar}
\end{figure}

In \cref{fig:diphoton_nnlo_qqbar} we first consider the  $q\bar{q}+q q'$ channels. 
Note that this channel doesn't receive $\PTB{}_\gamma$ improvements since there is only a gluon in the final state for the real-virtual contribution. 
We observe that the corrections from \PTB{} in $\tau_0$ and recoil in $q_T$ are substantial and 
allow asymptotic behavior to be reached much faster. This is in line with previous observations for
$q_T$ subtractions in \MATRIX \cite{Buonocore:2021tke}.

\begin{figure}
	\centering
	\includegraphics[width=0.49\columnwidth]{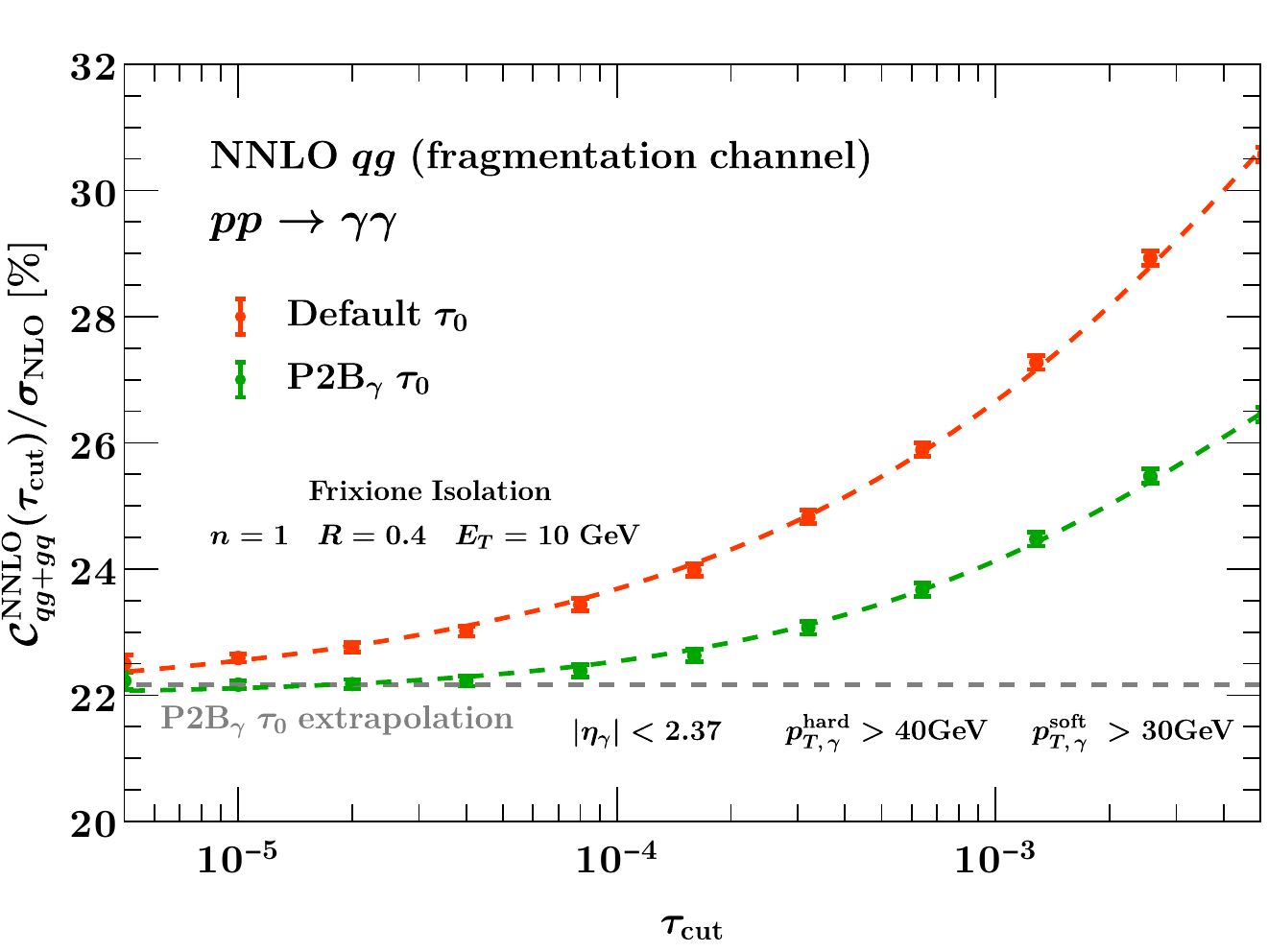}
	\includegraphics[width=0.49\columnwidth]{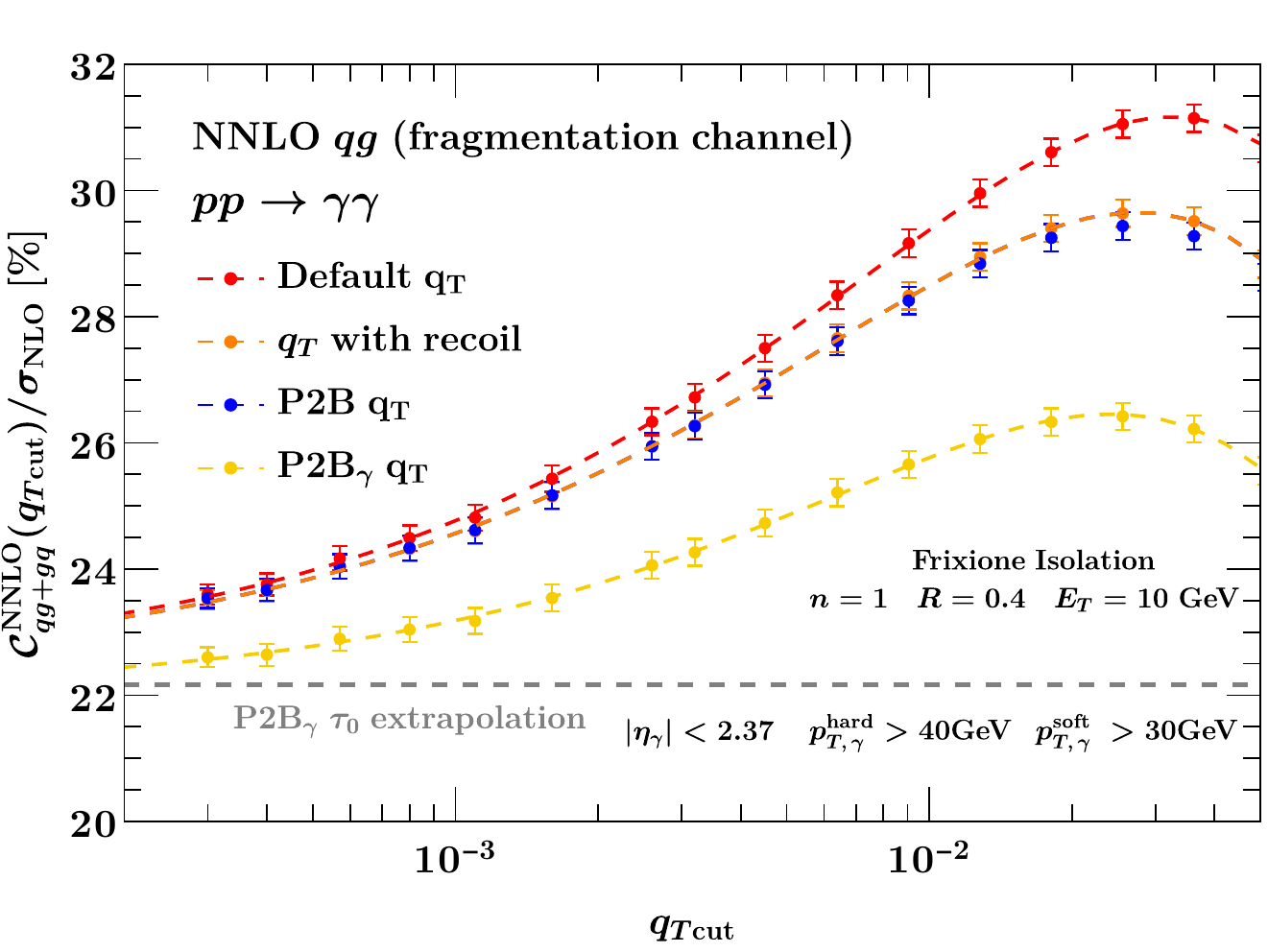}
	\caption{\NNLO{} coefficient K-factor for di-photon production, $qg$ fragmentation channel 
	only, for 
	$0$-jettiness (left) and $q_T$ subtractions (right) and their improved versions.}
	\label{fig:diphoton_nnlo_qg}
\end{figure}

In \cref{fig:diphoton_nnlo_qg} we consider the $qg$ fragmentation channels.
In this channel a quark in the final state is always present and an isolation procedure is required to define a finite cross-section. 
Here we distinguish between \PTB{} improvements and $\PTB{}_\gamma$ improvements. 
Looking first at the $q_T$ subtraction results in \cref{fig:diphoton_nnlo_qg}~(right) we see numerically equal improvements from the recoil and \PTB{} 
power corrections, as expected. Unlike for the Born channel, they are small and do not help noticeably. 
This issue is again in line with the findings in the literature \cite{Buonocore:2021tke}.
No reliable result can be obtained, even with power corrections from recoil, in stark contrast to the 
cases of $Z$ and $H$ production.

The power corrections associated with the photon isolation itself are dominant, and these are not captured by a recoil or \PTB{} approach.
On the other hand the $\PTB{}_\gamma$ improvements seen in \cref{fig:diphoton_nnlo_qg} are substantial. 
It stands out that, compared to the non-fragmentation channel, the $\tau_0$ subtractions perform 
substantially better than $q_T$ subtractions, reaching asymptotics already for $\tauc<10^{-4}$ down 
to the smallest value $5\cdot10^{-6}$ that can be achieved in double precision numerics. 
The huge improvements comparing the nominal slicing procedure to the new $\PTB{}_\gamma$ scheme are the same for $q_T$ and $\tau_0$ subtractions. 
For both slicing variables the $\PTB{}_\gamma$ procedure includes a sizable class of $\rcut$-linear power corrections. 

\begin{figure}
	\centering
	\includegraphics[width=0.7\columnwidth]{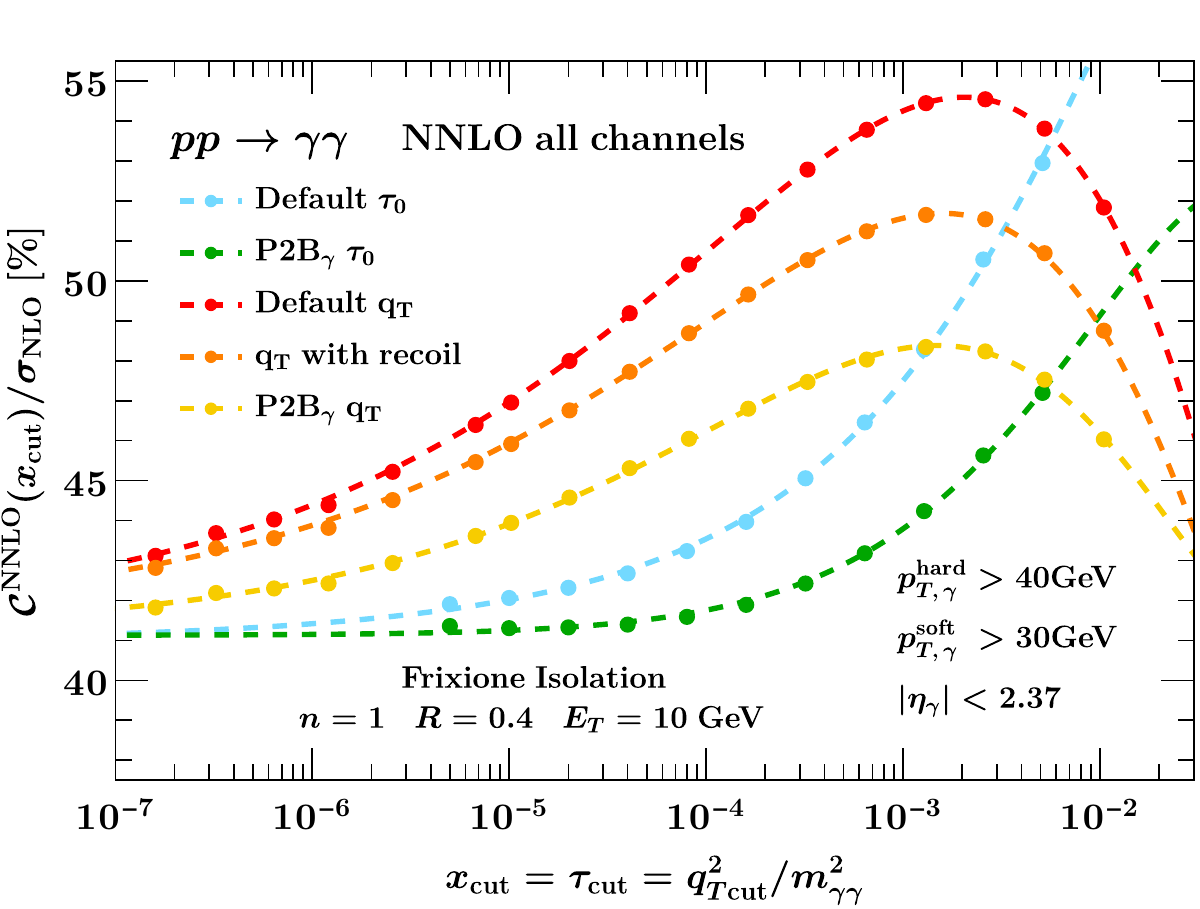}
	\caption{\NNLO{} coefficient K-factor for di-photon production for $0$-jettiness and $q_T$ 
	subtractions and their improved versions. Asymptotic fits take into account two to three 
	subleading terms. }
	\label{fig:diphoton_nnlo_sum}
\end{figure}

Finally, the sum of all channels (including the $gg$ component) is presented in 
\cref{fig:diphoton_nnlo_sum}, where we have used a fitted asymptotic extrapolation to continue the $q\bar{q}+qq'$ channel down to $\tcut=5\cdot10^{-6}$ for 
$0$-jettiness (it is already asymptotically flat at $\tauc=10^{-3}$, as can be seen in 
\cref{fig:diphoton_nnlo_qqbar}).
We performed asymptotic fits based on the two to three dominant subleading terms of the expected 
power corrections as laid out in \cref{sec:diphotonpc}. 
For $q_T$ subtractions with or without recoil improvements we observe that the asymptotic extrapolation has large uncertainties given that the distribution has not plateaued. On the other hand a robust asymptotic extraction is possible using $\PTB{}_\gamma+q_T$ and $0$-jettiness subtractions. 
In our study, the most reliable result is achieved by adding the $\PTB{}_\gamma$ corrections to 
$0$-jettiness subtractions. 
There one additionally reaps the benefit in the fragmentation channel of a better intrinsic scaling compared to $q_T$. 
The results achievable in double precision reach well into the asymptotic flat region, such that the constant piece in the asymptotic fit is well constrained. 

\section{Conclusions}\label{sec:conc}
The \LHC{} has enabled the measurement of many Standard Model processes with a high precision, 
challenging theoretical predictions that rely on continued developments in 
higher-order calculations. The continuation of this program with the {\abbrev HL-LHC} will 
significantly raise these challenges for the whole high-energy collider physics community, but is expected to result in significant scientific, 
technological and societal advancements.

One particularly important aspect of theoretical predictions is the combination of individually 
infrared divergent higher-order amplitudes into a finite cross-section that can be compared with 
experimental measurements. These subtraction methods are developed order by order in perturbation 
theory, and several different methods exist with individual benefits and drawbacks. 

In this paper we have introduced \PTB{}-improved $0$-jettiness and $q_T$ subtraction 
methods in the code \MCFM{}. 
The \PTB{}-improved scheme \cite{Ebert:2019zkb} includes power corrections that arise in the presence of fiducial cuts on the color-singlet system and are typically dominant compared to 
other subleading-power corrections. 
With these power corrections accounted for, much more reliable and robust results can be obtained. 
We exemplified the implementation for $Z$-boson production in the case of symmetric lepton cuts, and for Higgs production. 
Both for total cross-section rates as well as for rapidity distributions we find dramatic improvements in the processes examined, both at \NLO{} and \NNLO{}.
We find that the \PTB{}-$q_T$ method captures the same $\cO(q_{T,\text{cut}})$ power corrections as the recoil prescription for these Drell-Yan-type 
processes.
In the case of $0$-jettiness the \PTB{}-improvements allow one to additionally benefit, in the presence of realistic experimental cuts, from the analytically computed leading-logarithmic hadronic power corrections \cite{Moult:2016fqy,Moult:2017jsg,Ebert:2018lzn}, further improving the calculation of predictions.

For processes with photon isolation an additional source of power corrections arises due to the isolation procedure. 
These power corrections are typically large and dominant among the fiducial power corrections. 
In the non-fragmentation channel we find large improvements with both \PTB{} schemes, and also the recoil scheme in the case of $q_T$ subtractions.
In the numerically dominant fragmentation channel, where the collinear photon-quark singularity is removed, we find that the \PTB{} corrections (and recoil in the case of $q_T$ subtractions) are ineffective. 
The dominant set of power corrections is not removed by this procedure. 
In the case of $q_T$ subtractions with a recoil this can be readily understood, since the power correction procedure
does not take into account any information from the isolation prescription itself.

For this numerically-dominant fragmentation channel we developed a method, 
$\PTB{}_\mathbf{\gamma}$, to include the bulk of photon-isolation power corrections.
We numerically studied this for di-photon production, finding that it
significantly improved the calculation and 
enabled reliable and robust results in the presence of photon isolation. 
Additionally, we found that in the fragmentation channel the isolation power corrections of $0$-jettiness are substantially better than $q_T$-subtractions (compared to the non-fragmentation channels). 
The combination of intrinsically better power corrections and the photon-isolation-improved 
$\PTB{}_\gamma$ scheme allows us to extract a reliable result even without the need for extrapolation. 
We demonstrate asymptotically flat behavior for over an order of magnitude in the cutoff parameter. 
Without these improvements a significant cutoff dependence is present even for very small values of the cutoff, making the
extraction of a result with an error less than a few percent challenging to obtain.

The $\PTB{}$ and $\PTB_\gamma$ power corrections presented in this paper will be made available in 
the upcoming 10.4 release of \MCFM{} and complement and improve upon the existing recoil power 
corrections for $q_T$ subtractions.

\begin{acknowledgments}
We would like to thank Thomas Becher and Alexander Huss for discussion.
G.V. would like to thank Luca Buonocore for discussion and for providing the numbers for the comparison with \MATRIX \cite{Grazzini:2017mhc,Buonocore:2021tke} in \tab{MATRIXcomparison}. 
Tobias Neumann was supported by the United States Department of Energy under Grant Contract 
DE-SC0023047, \enquote{The three-dimensional structure of the proton}. This research used resources 
of the National Energy Research Scientific Computing 
Center (NERSC), a 
U.S. Department of Energy Office of Science User Facility located at Lawrence Berkeley National 
Laboratory, operated under Contract No. DE-AC02-05CH11231 using NERSC award HEP-ERCAP0023824 
\enquote{Higher-order calculations for precision collider phenomenology}.
This research was supported by the Fermi National Accelerator
Laboratory (Fermilab), a U.S.\ Department of Energy, Office of
Science, HEP User Facility managed by Fermi Research Alliance,
LLC (FRA), acting under Contract No. DE--AC02--07CH11359.

\end{acknowledgments}

\bibliography{p2bbib}{}
\bibliographystyle{jhep}

\end{document}